\documentclass[twoside,a4paper]{article}
\usepackage{helvet,times}
\usepackage{bm,textcomp}
\usepackage{url}
\usepackage{subcaption}
\usepackage{tikz}
\usepackage{floatrow}
\usepackage{lipsum}
\usepackage{wrapfig}
\usepackage[innercaption]{sidecap}    
\sidecaptionvpos{figure}{t}
\usepackage{xr}
\usepackage{geometry}
\newgeometry{vmargin={15mm}, hmargin={20mm,20mm}}
\makeatletter
\newcommand*{\addFileDependency}[1]{
  \typeout{(#1)}
  \@addtofilelist{#1}
  \IfFileExists{#1}{}{\typeout{No file #1.}}
}
\makeatother
 
\newcommand*{\myexternaldocument}[1]{%
    \externaldocument{#1}%
    \addFileDependency{#1.tex}%
    \addFileDependency{#1.aux}%
}

\myexternaldocument{supplementary}

\usepackage{amsmath}

\usepackage[round,numbers,sort&compress]{natbib}

\begin{document}

\setcounter{page}{1} 

{\fontsize{17}{12}\selectfont
Stochastic model of T Cell repolarization during target elimination (I)
}\\

\begin{center}
{\fontsize{12}{12}\selectfont
Ivan Hornak$^{1,*}$, Heiko Rieger$^{1}$\\
\vspace{1mm}
$^{1}$Department of Theoretical Physics and Center for Biophysics, Saarland University,
Saarbr{\"u}cken 66123, Germany\\
\vspace{1mm}
$^{*}$hornak@lusi.uni-sb.de
}\\
\end{center}

\section*{ Abstract}
Cytotoxic T lymphocytes (T) and natural killer (NK) cells are the main cytotoxic killer cells of the human body to eliminate pathogen-infected or tumorigenic cells (i.e. target cells). Once a NK or T cell has identified a target cell, they form a tight contact zone, the immunological synapse (IS). One then observes a re-polarization of the cell involving the rotation of the microtubule (MT) cytoskeleton
 and a movement of the microtubule organizing center (MTOC) to a position that is just underneath the plasma membrane at the center of the IS.  
 Concomitantly a massive relocation of organelles attached to MTs is observed, including the Golgi apparatus, lytic granules and mitochondria. Since the mechanism of this relocation is still elusive we devise a theoretical model for the molecular motor driven motion of the MT cytoskeleton
 confined between plasma membrane and nucleus during T cell polarization. We analyze different scenarios currently discussed in the literature, the cortical sliding and the capture-shrinkage mechanisms, and compare quantitative predictions about the spatio-temporal evolution of MTOC position and MT cytoskeleton
  morphology with experimental observations. The model predicts the experimentally observed biphasic nature of the repositioning due to an interplay between MT cytoskeleton
  geometry and motor forces and confirms the dominance of the capture-shrinkage over the cortical sliding mechanism when MTOC and IS are initially diametrically opposed. We also find that the two mechanisms act 
synergistically,  
   thereby reducing the resources necessary for repositioning. Moreover, it turns out that the localization of dyneins in the pSMAC facilitates their interaction with the MTs. Our model also opens a way to infer details of the dynein distribution from the experimentally observed features of the MT cytoskeleton
  dynamics.  In a subsequent publication we will address the issue of general initial configurations and situations in which the T cell established two immunological synapses.

\section{Introduction}
\label{introduction}

Cytotoxic T lymphocytes and  natural killer cells have a key role in our immune system by 
finding and destruction of virus-infected and tumor cells, parasites and foreign invaders.
Once a T cell leaves the thymus, it circulates through the organism in search of 
a target cell.
The directional killing of a target cell
 is completed in three subsequent steps.
First, T cell receptors (TCR) bind to antigens on the surface of the 
target cell
 presented by the major histocompatibility complex 
\cite{rudolph_how_2006,garcia_reconciling_2012, zinkernagel_restriction_1974,attaf_t_2015,wucherpfennig_t_2004,babbitt_binding_1985}
 leading to the creation of a tight contact 
zone called immunological synapse (IS)
\citep{monks_three-dimensional_1998,dustin_novel_1998,dustin_understanding_2010}
composed of multiple supramolecular activation clusters 
\cite{dustin_understanding_2010,huang_deficiency_2005,andre_use_1990}. 
In the second step, the cell repolarizes by relocating the 
microtubule organizing center (MTOC) towards the IS \cite{geiger_spatial_1982,kupfer_polarization_1982,yi_centrosome_2013,
stinchcombe_centrosome_2006,maccari_cytoskeleton_2016,kuhn_dynamic_2002,hui_dynamic_2017} under the influence of forces exerted on MTs.
Moreover, since mitochondria, Golgi apparatus and Endoplasmic reticulum  are attached to MTs, these organelles are dragged along with the MT cytoskeleton
 \cite{maccari_cytoskeleton_2016,kupfer_reorientation_1984,kupfer_specific_1986,gurel_connecting_2014,lee_dynamic_1988,waterman-storer_endoplasmic_1998,palmer_role_2005}. 
Consequently, the repolarization process involves massive rearrangements of the internal, MT associated structure of the cell.
In the third step, the T cell releases at the IS the
cytotoxic material (e.g. the pore forming protein perforin and 
the apoptosis inducing granzyme) from vesicles, the lytic granules, \cite{mullbacher_granzymes_1999,lowin_perforin_1995,voskoboinik_perforin-mediated_2006,grossman_orphan_2003,krzewski_human_2012} 
towards the  target cell
leading to its destruction
  \cite{krzewski_human_2012,yannelli_reorientation_1986,pasternack_serine_1986,poo_receptor-directed_1988,kupfer_small_1994,stinchcombe_immunological_2001,
  haddad_defective_2001,griffiths_protein_1997,stinchcombe_rab27a_2001,calvo_imaging_2018,kupfer_reorientation_1985}.
Although the lytic granule secretion can take place
without the MTOC repolarization \cite{golstein_early_2018}, or before it \cite{bertrand_initial_2013}, the MTOC accompanied granule secretion may be required for the killing of resistant cells, such as tumor cells.

The IS is partitioned into several supramolecular activation clusters (SMACs) including the distal SMAC (dSMAC), peripheral SMAC (pSMAC) and the central SMAC (cSMAC) \cite{monks_three-dimensional_1998,dustin_understanding_2010,andre_use_1990,lin_c-smac_2005,choudhuri_signaling_2010}, 
in which TCR (cSMAC) and adhesion molecules are organized.
Moreover, the engagement with the target cell results in the formation of actin and actomyosin networks at the IS \cite{hammer_origin_2019}. 
Dynein, a minus-end directed (towards the MTOC) molecular motor protein
anchored at the cell cortex, is absolutely necessary for the repolarization to take place as was experimentally demonstrated with knock-out experiments  
\cite{martin-cofreces_mtoc_2008},
analogous to dynein exerting forces against anchor proteins fixed 
at the cell cortex during mitosis\cite{nguyen-ngoc_coupling_2007,saito_mcp5_2006,yamashita_fission_2006,ananthanarayanan_dynein_2013}.

Once the T cell is activated, the adaptor protein ADAP forms a ring at the periphery of the IS, with which dynein colocalizes \cite{combs_recruitment_2006,hashimoto-tane_dynein-driven_2011}.
Concerning the underlying mechanism, it was proposed that the repolarization is driven by the cortical sliding mechanism  \cite{combs_recruitment_2006,stinchcombe_communication_2014},
in which dyneins 
anchored at the cell cortex 
step on the MT towards the minus-end and thus pull 
the MTOC towards the IS.
The first experimental indications for the cortical sliding came from the observation of the cytoskeleton movement using the polarization light microscopy \cite{kuhn_dynamic_2002}.  
Subsequent experiments indicate that the IS periphery, in particular the ring shaped pSMAC, is the region where dyneins attach to and pull on MTs \cite{combs_recruitment_2006,kuhn_dynamic_2002}.

The repositioning was observed in various experiments.
Focused activation of the photo-activable peptide-MHC on the glass surface was used in \cite{quann_localized_2009}.
In \cite{maccari_cytoskeleton_2016} the repositioning was
observed alongside with the rotation of the mitochondria, which provided evidence that the mitochondria are dragged with the MT cytoskeleton.
Detailed observations were made by Yi et al 
\cite{yi_centrosome_2013} providing a new insight into the mechanism of
the repolarization.
In \cite{yi_centrosome_2013} an optical trap was used to place 
a target cell
so that the initial point of contact is in diametrical opposition to  the current position of the MTOC, which allowed for the dynamical imagining in a quantitative fashion.
During the experiment, the deformations and changes in MT structures were observed and the position of the MTOC tracked.
First of all, Yi et al. \cite{yi_centrosome_2013} provided  strong experimental evidence against the cortical sliding mechanism. Instead, the 
observations indicate that the MTOC is driven by a capture-shrinkage mechanism  \cite{laan_cortical_2012} localized in a narrow central region 
of the IS.
The capture-shrinkage mechanism involves dynein interacting in an end-on fashion with the plus-end of a MT, which is fixed in a position on the membrane 
of the cell where the MT depolymerizes.
The MT shrinkage part happens plausibly because dynein pulls the MT plus-end against the cell membrane, which increases the force dependent MT depolymerization rate \cite{laan_cortical_2012}.

In sequences of microscope pictures \cite{yi_centrosome_2013} showed that 
MTs reach from the MTOC to the IS and bend alongside the cell membrane.
Subsequently, MTs form a narrow stalk connecting the MTOC with the center of the IS. 
The plus-end of MTs in the stalk, while captured in the center of the IS, straighten (probably under tension due to the dynein pulling at the plus-end) and shrink by depolymerization at the capture point. 
Consequently, the MTOC is dragged towards the center of IS, which invaginates into the cell, further proving the 
location of the main pulling force.
When the MT depolymerization was inhibited by taxol, 
the MTOC  repositioning slowed down substantially.  
These observations supported the hypothesis that the capture-shrinkage mechanism plays a major role.
However, the velocity of the MTOC repositioning is not always the advantage, since time is necessarily for the killing of target cells in hostile environments \cite{breart_two-photon_2008} and might be beneficial for slower killing processes \cite{he_ctls_2007}. 
Additionally, Yi et al \cite{yi_centrosome_2013} reported that the repositioning is biphasic and that the two phases differ in the velocity of the MTOC and the orientation of its movement.
In the first, so-called polarization phase, the MTOC travels quickly around the nucleus of the cell in a circular manner. 
The polarization phase ends when the MTOC is approximately $2\mu\textrm{m}$ from the center of the IS.
Subsequently, during the second "docking phase" the MTOC travels directly 
towards the IS with a substantially decreased velocity.

The cortical sliding mechanism alone was previously analyzed with a deterministic mechanical model \cite{kim_deterministic_2009}, where it was demonstrated that mechanism is capable of reorienting the MTOC into a position under the IS underneath certain conditions.
Furthermore, oscillations between two IS were studied in different situations. 
Nevertheless, 
the forces in the model were deterministic, neglecting the stochastic nature of dynein attachment, detachment and stepping, leaving various experimental observations unexplained, as for instance the preferential attachment of MTs to a dynein anchored in the periphery of the IS.

Sakar et al \cite{sarkar_search_2019} hypothesized that dynamic MTs find the central region of the IS, where they can be captured by a dynein by growing from 
the MTOC in random directions, analogous to the search and capture mechanism during the formation of the mitotic MT cytoskeleton.
Once MTs attach to the dynein 
in the central region of the IS, the relocation of the MTOC starts, which is
the process that is analyzed in this paper.

In spite of these detailed experimental observations, many aspects of the internal mechanisms driving the relocation of the MTOC during the T cell repolarization  remain poorly understood,
like the cause of the transition from the polarization to the docking phase.
Yi et al argue that a resistive force emerges when the MTOC-IS distance is around $2\mu\textrm{m}$ leading to a reduction in the MTOC velocity.
The potential causes are physical impediments to the MTOC translation or a reduced attachment or a force development of molecular motors.
Moreover, the experiments of Yi et al were performed with specific initial positions of the IS and the MTOC, being diametrically opposed.
The question arises whether the observed dominance of the capture-shrinkage mechanism would be robust in other naturally occurring situations in which the initial position of the MTOC is not in the diametrical opposition to the IS.
If the capture-shrinkage is the truly dominant mechanism, what is the role of the cortical sliding?
Finally, why are cortically sliding MTs  caught just on the periphery of  the IS \cite{kuhn_dynamic_2002}, is it caused purely by the colocalization of dyneins with the ADAP ring \cite{combs_recruitment_2006}? 
The answers to these questions are still elusive and in this work we analyze them in the framework of a quantitative theoretical model for the relocation of the MTOC after the IS formation.
Although this study focuses on the T Cell,
NK cells display the same kind of phenomenology: the IS formation, the MTOC relocation, the release of lytic granules.

We distribute our analysis into two consecutive publications. In this, first, publication we describe the theoretical model we use and present our results focusing on the experiments described in \cite{yi_centrosome_2013}, \cite{kuhn_dynamic_2002}
and \cite{combs_recruitment_2006} and on an analysis of the two mechanisms: the cortical sliding and the capture shrinkage. This comprises the setup in which the T cell has one IS and the initial positions of the IS and the MTOC are diametrically opposed to each other.

A subsequent, second, publication will focus on quantitative predictions of our model for situations that have not yet been analyzed experimentally. There we will focus on the repolarization following initial configurations not realized in (43), which will also provide additional insight into the different effects of the two mechanisms, the cortical sliding and the capture shrinkage. Moreover we will analyze the, eventually oscillating, MT/MTOC movement 
with two IS.

\section{Methods}
\subsection{Computational model}

\begin{figure*}[ht]
     \includegraphics*[trim=50 535 50 68,clip,width=6.75in]{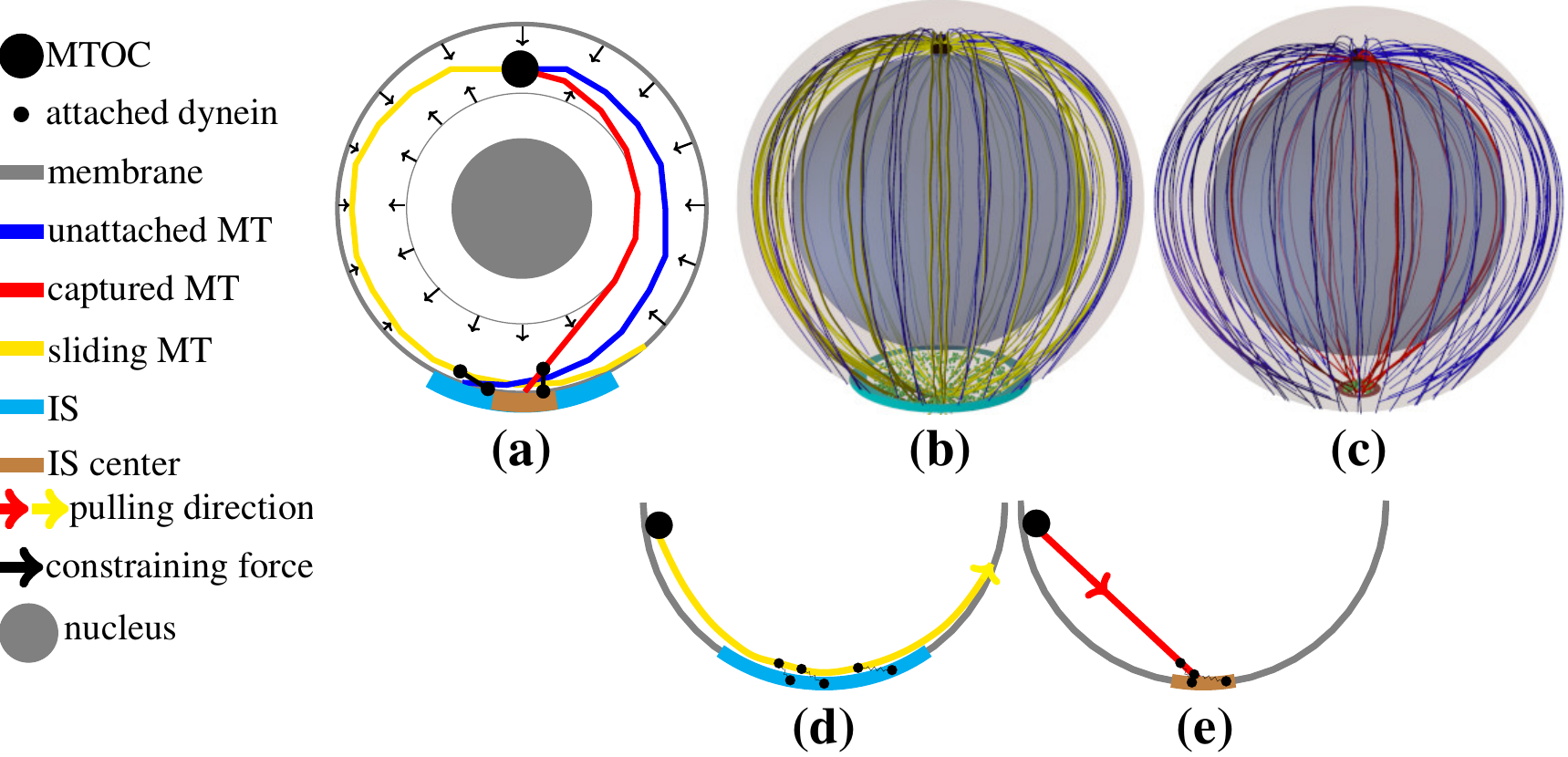} 
   \caption{\small{
(a)(b)(c)  Sketch of the model.
(a)Two-dimensional cross-section of the model. 
MTs sprout from the MTOC and their movement is confined by constraining forces from the cell membrane and the nucleus.  
MTs are attached to dynein motors in the IS and they are pulled by the capture-shrinkage or the cortical sliding mechanism. 
(b)(c)Three-dimensional sketch of of the cell model.
The outer transparent and inner spheres represent the cell membrane and the nucleus of the cell, respectively.
(b)
Blue disk represents the IS, where cortical sliding dynein is anchored.
Small green dots in the IS represent randomly distributed dynein.
(c) Brown disk represents the central region of the IS where the capture-shrinkage dynein is anchored.  
(d)(e) Sketch of the cortical sliding mechanism (d) and the capture-shrinkage
mechanism (e). Small black dots on the membrane: dynein anchor points, small black dots on the MTs: attachment points. Note that MT depolymerize when 
pulled by capture-shrinkage dynein towards the membrane.
       \label{cell} }}   
 \end{figure*}

\small{
The cell and its nucleus are modeled as two concentric spheres of the radius $5\mu\textrm{m}$ and $3.8\mu\textrm{m}$, respectively.
The model of the cytoskeleton consists of microtubules (MT) and the microtubule organizing center (MTOC), see Fig. \ref{cell}. 
MTs are thin filaments with a diameter of ca $25\textrm{nm}$ 
\cite{cooper_cell_2000,li_microtubule_2002,meurer-grob_microtubule_2001}. 
The measured values of the flexural rigidity varies between experiments
\cite{jia_measuring_2017,takasone_flexural_2002}, 
in our model we take $2.2 *10^{-23} \textrm{N} \textrm{m}^{2}$ 
\cite{gittes_flexural_1993} yielding a persistence length larger than  $5 \textrm{mm}$
exceeding the size of the cell by three orders of magnitude.
A single MT is represented by a bead-rod model \cite{broedersz_modeling_2014}.
Since repolarization occurs on a time scale of seconds, the growth of MTs is neglected.
The beads move under the influence of forces to be described below (and defined in detail in the SI \ref{micro_discretization}(a): bending, drag, molecular motor and stochastic forces). Assuming zero longitudinal elasticity of the MTs, we use constrained Langevin dynamics to model the motion of the MTs, see SI.
Repulsive forces acting on the MT segments confine the cytoskeleton between the nucleus and the cell membrane.
}

\small{
The MT organizing center (MTOC) is a large protein complex which has a complex structure 
composing of mother and daughter centrioles \cite{chretien_reconstruction_1997,winey_centriole_2014,guichard_native_2013,bernhard_[electron_1956}
 embedded 
in the  pericentriolar material(PCM) \cite{woodruff_pericentriolar_2014,moritz_structure_2000,robbins_centriole_1968}.
MTs nucleate from gamma-tubulin-containing ring structures 
 within the PCM mainly at the appendages of the mother centriole
\cite{moritz_microtubule_1995,chretien_reconstruction_1997}. 
MT can sprout from the MTOC in all directions. 
MTs 
whose original direction is approximately parallel to the membrane of the cell, will continue to grow to the cell periphery.
Other MTs will soon hit the wall of the cell or its nucleus. 
Such MT can either bend and assume a new direction parallel to the cell membrane, 
or undergo the MT catastrophe.
Therefore, the long MTs are seemingly always sprouting from the MTOC in one plane, as can be seen in \cite{yi_centrosome_2013}.
Consequently,  we model the MTOC as a planar, rigid,  polygon structure (Fig. S\ref{MTOC_scatch} in the Supporting Material) from which MTs emanate in random directions by fixing the positions and directions of their first segment, (Fig. S\ref{MTOC_scatch_2}).
MTs sprout from the MTOC to the cell periphery, see \ref{cell}a.
}

\small{
Unattached dynein is represented just with one point on the surface of the cell.
If the dynein is closer to the MT than $L_{0}$,
protein attaches with a probability $p_{a}$.
Dynein motors are distributed randomly in specific, spatially varying concentrations, on the cell boundary. Attached dynein is represented by a fixed anchor point located on the cell boundary and an attachment point located on a MT, both being connected by an elastic stalk of a length $L_{0}$ \cite{burgess_dynein_2003,belyy_mammalian_2016}.
The force exerted on a MT $F^{\textrm{Dynein}}_{i}$ depends on the stalks elastic modulus $k_{\textrm{Dynein}}$ and its prolongation.
The dynein stepping depends on the magnitude of the force and its orientation.
If the force is parallel to the preferred direction of the  stepping, the attachment point moves one step to the MT minus-end (towards the MTOC) with a constant probability $p_{-}$.
If the orientation of the force is opposite and its magnitude smaller than a stall force $F_{S}$, dynein makes one step towards the minus-end with a force-depending probability.
If $|F^{\textrm{Dynein}}_{i}| > F_{S}$ and the force
has a unfavorable direction,  the 
dynein makes one step to the plus-end with a constant probability $p_{+}$. 
The steps of the dynein have varying lengths\cite{belyy_mammalian_2016} but for simplicity we set it to the most frequently measured value of $d_{\textrm{step}}=8n\textrm{m}$.
The probability of detachment,
$p_{\textrm{detach}}$, increases with the force.

Experimentally, two mechanisms by which the dynein act on MTs have been identified: the cortical sliding \cite{kuhn_dynamic_2002}, where MTs under the effect of dynein move tangentially along the membrane, and the capture shrinkage \cite{yi_centrosome_2013}, by which MTs under the effect of dynein are reeled in towards the membrane and concomitantly depolymerized
 (sketched in Figs. \ref{cell}(d) and \ref{cell}(e)).
}

\small{
The IS is divided into two regions: the center, where dyneins act on MTs via the capture shrinkage mechanism \cite{yi_centrosome_2013} and the complete IS, where dyneins act via the cortical sliding mechanism. Each region is modeled as an intersection of the cell sphere with a cylinder, \ref{cell}, with radius $R_{\textrm{IS}} = 2 \mu \textrm{m}$ for the complete IS and $R_{\textrm{CIS}} = 0.4 \mu \textrm{m}$ for the central region. 
Dyneins are distributed randomly with uniform area density 
 $\rho_{\textrm{IS}}$ in the small central region, denoted as capture-shrinkage dynein, and density $\tilde{\rho}_{\textrm{IS}}$ in
 the larger region of the whole IS, denoted as cortical-sliding dynein.

\section{Results}
\normalsize

We analyzed the role of the cortical sliding and the capture shrinkage mechanisms
and their combined effect
during the repolarization by computer simulations of the model defined in the previous section. 
The density of dyneins anchored at the IS, $\tilde{\rho}_{IS}$, and the central region of
the IS, $\rho_{IS}$ are unknown model parameters which we therefore 
vary over a broad range between 0 (no anchored dynein) and 1000$\mu \textrm{m}^{-2}$
(the maximum number of anchored dynein due to the lateral size of dyneins,
see SI 1.1.5).
During the integration of the equation of motion, various quantities are calculated: the distance between the center of the MTOC and the IS, $d_{\textrm{MIS}}$, the number of dyneins attached to the MTs, $N_{\textrm{dm}}$, the velocity of the MTOC, $v_{\textrm{MTOC}}$, the distance between the MTOC and the center of the cell, $d_{\textrm{MC}}$.
For each point in the parameter space these quantities were averaged over 500 simulation runs. 
Each simulation run is initialized with the mechanical equilibrium (minimum elastic energy) configuration of the MT/MTOC-system and all dyneins being detached. 
Results are shown with the standard deviation as error bars only when they are larger than the symbol size.

\subsection{Capture-Shrinkage mechanism}
\label{Capture_Shrinkage_mechanism}

\begin{figure*}[ht]
    \includegraphics*[trim=15 370 40 60,clip,width=6.75in]{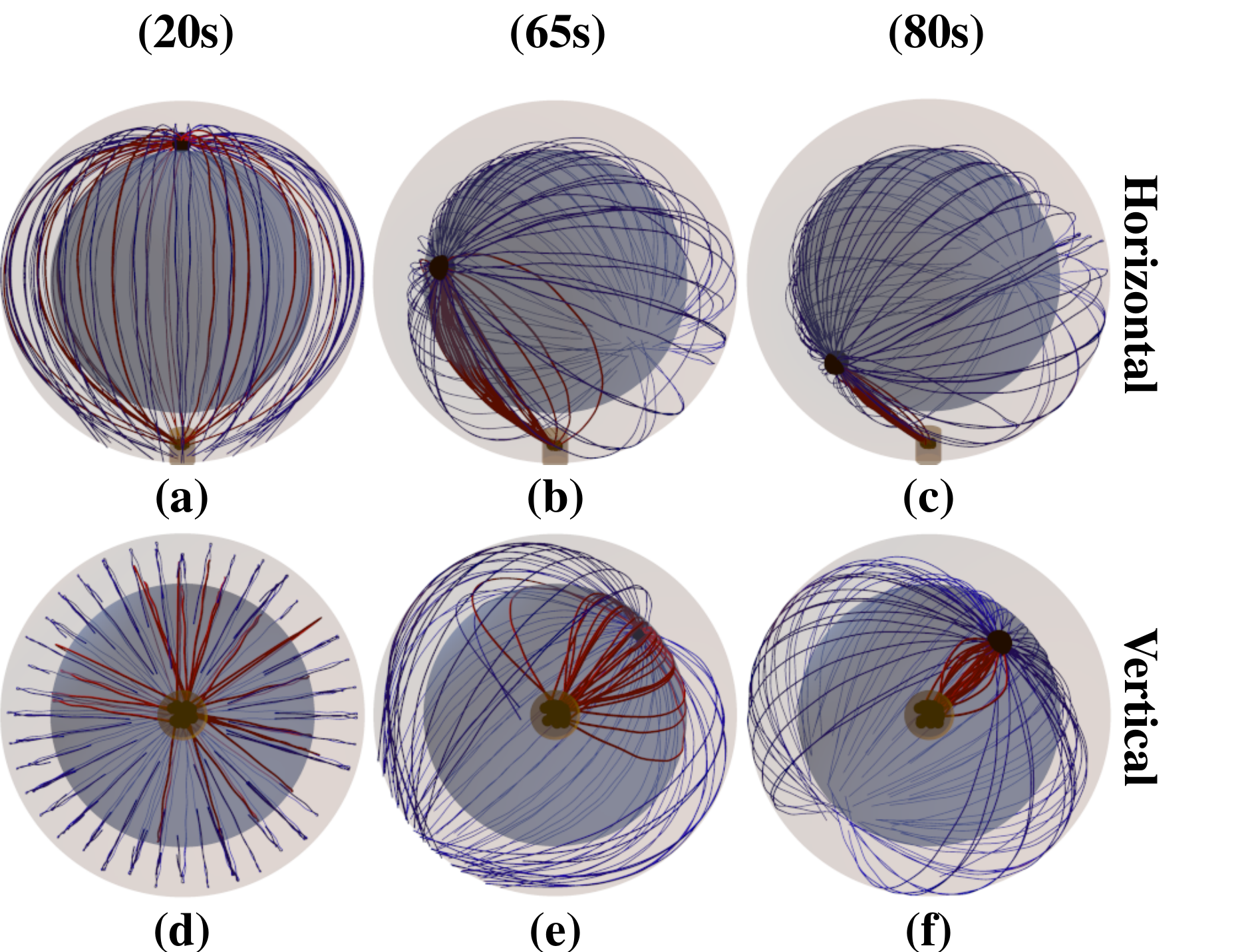}    
\caption{Snapshots from the time-evolution of the MT cytoskeleton
 configuration under the effect of the capture-shrinkage mechanism alone (dynein density 
$\rho_{\textrm{IS}} = 100\mu \textrm{m}^{-2}$).
MTs are connected to the MTOC indicated by the large black sphere.
Blue and red curves are unattached and attached MTs.
Small black spheres in the IS represent dyneins. 
The brown cylinder indicates the center of the IS, where the 
capture-shrinkage dyneins are located.
 (a)(d)$d_{\textrm{MIS}} = 9\mu\textrm{m}$. Initially the attached MTs sprout from the MTOC in all directions directions. (b)(e) $d_{\textrm{MIS}} = 6\mu\textrm{m}$. As time progresses, microtubules form a stalk connecting the MTOC and the IS. (c)(f) $d_{\textrm{MIS}} = 2.5\mu\textrm{m}$. The stalk is fully formed and it shortens as the MTOC approaches the IS.    
 \label{fig:opening_capt_shr}
}
 \end{figure*}

The repositioning process under the effect of the capture-shrinkage mechanism is visualized in Fig. \ref{fig:opening_capt_shr}.
In Figs. \ref{fig:opening_capt_shr}a and \ref{fig:opening_capt_shr}d, it can be seen that initially the attached MTs aim from the MTOC in all directions. Subsequently, the stalk of MTs is almost formed in the middle phase of
the repositioning, Fig. \ref{fig:opening_capt_shr}b and Fig. \ref{fig:opening_capt_shr}e, and it is fully formed as the MTOC approaches the IS, see Fig. \ref{fig:opening_capt_shr}c and \ref{fig:opening_capt_shr}f
and the supplementary Movie $\textrm{S1}$ showing
the time-evolution of the MT cytoskeleton
 configuration under the effect of the
capture-shrinkage mechanism with 100 MTs and the dynein density 
$\rho_{\textrm{IS}} = 100\mu \textrm{m}^{-2}$.

The process can be divided into three phases based on the time-evolution 
of the MTOC velocity, see Fig. \ref{fig:capture_shrinkage_1}b.  
In the first phase, when the distance between the MTOC and the center of the IS
is $\bar{d}_{\textrm{MIS}}>8.8\mu\textrm{m}$, the velocity changes
rapidly in the first seconds of the process and then falls to a local minimum.
In the second phase, the velocity continually increases to a maximum and then in the third phase, it decreases again. 
By comparison of Fig. \ref{fig:capture_shrinkage_1}b and Fig. \ref{fig:capture_shrinkage_1}c,
it can can be seen that the time-evolution of the velocity corresponds to the time-evolution of the number of dyneins acting on MTs.
The evolution of the number of attached dyneins during the first phase can be understood from an analysis of Figs. \ref{fig:capture_shrinkage_1}d,
\ref{fig:opening_capt_shr}a and \ref{fig:opening_capt_shr}b. 
At the beginning of the simulations, a substantial number of MTs intersects 
the IS (visually demonstrated in Fig. \ref{fig:opening_capt_shr}a and Fig. \ref{fig:opening_capt_shr}d) resulting in a fast increase of the number of attached dyneins.
Since the MTs attached to dynein sprout from the MTOC in every direction,
c.f. Fig. \ref{fig:capture_shrinkage_1}e, the MTOC moves towards the IS and, simultaneously, to the nucleus of the cell, see Fig. \ref{fig:capture_shrinkage_1}d.
As the MTOC approaches the nucleus of the cell, the nucleus starts to oppose 
the movement by repelling the MTs and, at the end of the first phase, the MTOC.
Therefore, as the pulling force of the dyneins is opposed by the nucleus, the dyneins detach since the detachment rate is force-dependent.

\begin{figure*}
     \includegraphics*[trim=45 495 50 70,clip,width=6.75in]{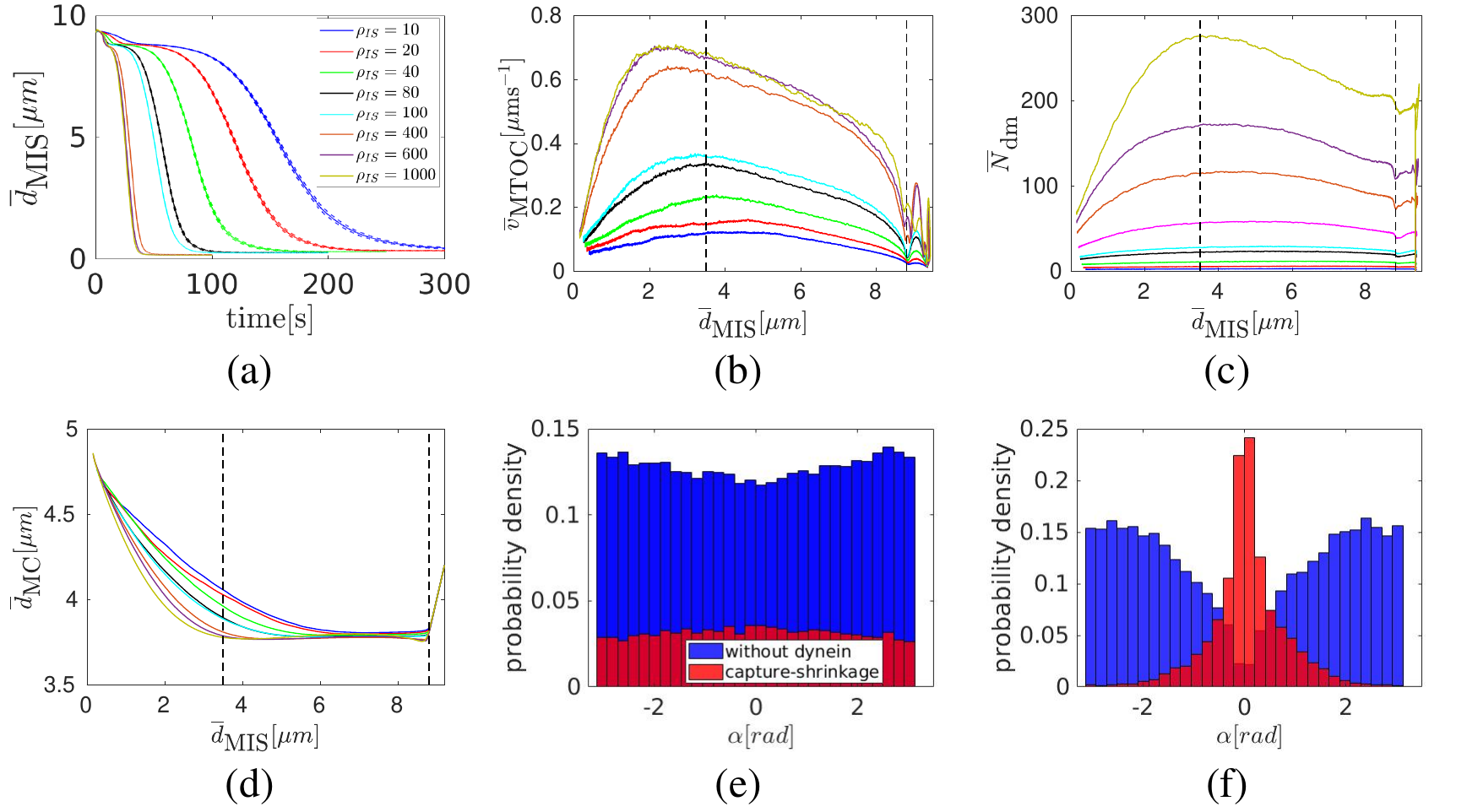}    
      \caption{Capture-shrinkage mechanism:
(a) The dependence of the average MTOC-IS distance $\bar{d}_{\textrm{MIS}}$ on time.
(b)(c)(d) Dependencies  of the average MTOC velocity $\bar{v}_{\textrm{MTOC}}$(b), the number of dyneins acting on microtubules $\bar{N}_{\textrm{dm}}$(c) and the
MTOC-center distance $\bar{d}_{\textrm{MC}}$(d) on the average MTOC-IS distance. Black dashed lines denote transitions between different phases
of the repositioning process.
(e)(f)    
Probability distributions
 of the angles between the first MT segments  and the direction of the MTOC movement for a dynein density $\rho_{\textrm{IS}} = 100 \mu\textrm{m}^{-2}$
(e) $t = 1\textrm{s}$, $\bar{d}_{\textrm{MIS}} \sim 9\mu\textrm{m}$.
(f) $t = 60\textrm{s}$, $\bar{d}_{\textrm{MIS}} \sim 5\mu\textrm{m}$. 
       \label{fig:capture_shrinkage_1}}
 \end{figure*}

The increase of the number of attached dyneins $\bar{N}_{\textrm{dm}}$ in the second phase 
can be explained by considering the fact that the MTOC slides over 
the surface of the nucleus and the MT stalk forms.
At the beginning, the nucleus presents an obstacle between the MTOC and 
the IS, see Fig. \ref{fig:opening_capt_shr}a.
The opposing force from the nucleus decreases with the approach of  the MTOC towards the IS.
At the end of the repositioning, the nucleus does not longer stands between the two objects, see Fig. \ref{fig:opening_capt_shr}c and
\ref{fig:opening_capt_shr}f.
Therefore, the opposing force from the nucleus contributing to dynein detachment decreases.
More importantly, attached MTs form the MT stalk.
The angle $\alpha$ between the first segment of the MT and the direction of the MTOC movement is used to describe the deformation of the cytoskeleton structure and the stalk formation.
At the beginning of simulation(the first phase and the beginning of the second), attached MTs aim in every direction, see Fig.  \ref{fig:capture_shrinkage_1}e, visualized in Fig.
\ref{fig:opening_capt_shr}a and \ref{fig:opening_capt_shr}d.
Therefore, the dyneins pull in multiple directions which makes them oppose each other leading to dynein detachment.
After a few seconds, the MTOC travels in the direction of the
biggest pulling force. 
Consequently, the attached MTs form a stalk as the simulation progress  and dyneins act in alignment, see Figs. \ref{fig:capture_shrinkage_1}f,  \ref{fig:opening_capt_shr}b and \ref{fig:opening_capt_shr}e.  
They do not longer oppose each other, but they share the load from opposing forces.
Consequently, the detachment probability of dynein decreases with the opposing force and the number of attached dyneins increases.

The number of dyneins decreases in the final phase when $\bar{d}_{\textrm{MIS}}<3.5\mu\textrm{m}$, see Fig. \ref{fig:capture_shrinkage_1}c.
Unattached MTs in the IS are pushed backward by viscous drag  as the MTOC moves to the IS.
As a result, one observes an "opening" of the MT cytoskeleton, c.f. Figs.
\ref{fig:capture_shrinkage_1}e, \ref{fig:opening_capt_shr}c and \ref{fig:opening_capt_shr}f.
Unattached MTs do not intersect the IS, see 
Fig. \ref{fig:capture_shrinkage_1}f, and cannot attach to dyneins.
The attached MTs shorten due to the depolymerization further lowering the probability of dynein attachment.  
Moreover, an opposing force arises from the cytoskeleton being dragged from the nucleus to the membrane, see Fig. \ref{fig:capture_shrinkage_1}d,
causing the detachment of dyneins since the detachment rate is force dependent.

To summarize, the trajectory of the MTOC towards the IS displays 
three phases, where the two longer phases have been reported also in
the experiment \cite{yi_centrosome_2013} but not the short initial 
phase. First the MTOC descends to the nucleus, see Figs. \ref{fig:capture_shrinkage_1}a and \ref{fig:capture_shrinkage_1}d,
then it moves to the IS fast and then slows down during the last $2\mu\textrm{m}$, see Fig. \ref{fig:capture_shrinkage_1}b. 
Once the MTOC bypasses the nucleus, it moves away switching
from a purely circular to partially radial movement, see Fig. \ref{fig:capture_shrinkage_1}d.
The variation of the MTOC velocity, its modulus and its direction, 
is clearly visible in the supplementary Movie $\textrm{S2}$, 
showing a simulation with a smaller nucleus radius 
$r_{N} = 3.3\mu\textrm{m}$.
 Note that the duration of the 
complete repositioning process in the
experiments is ca 60-90 sec, which our model predicts to be achieved 
by a dynein density of 80-200 $\mu$m$^{-2}$.

\subsection{Cortical Sliding mechanism}
\label{Cortical_Sliding_mechanism_in_IS}

\begin{figure*}
     \includegraphics*[trim=15 370 40 60,clip,width=6.75in]{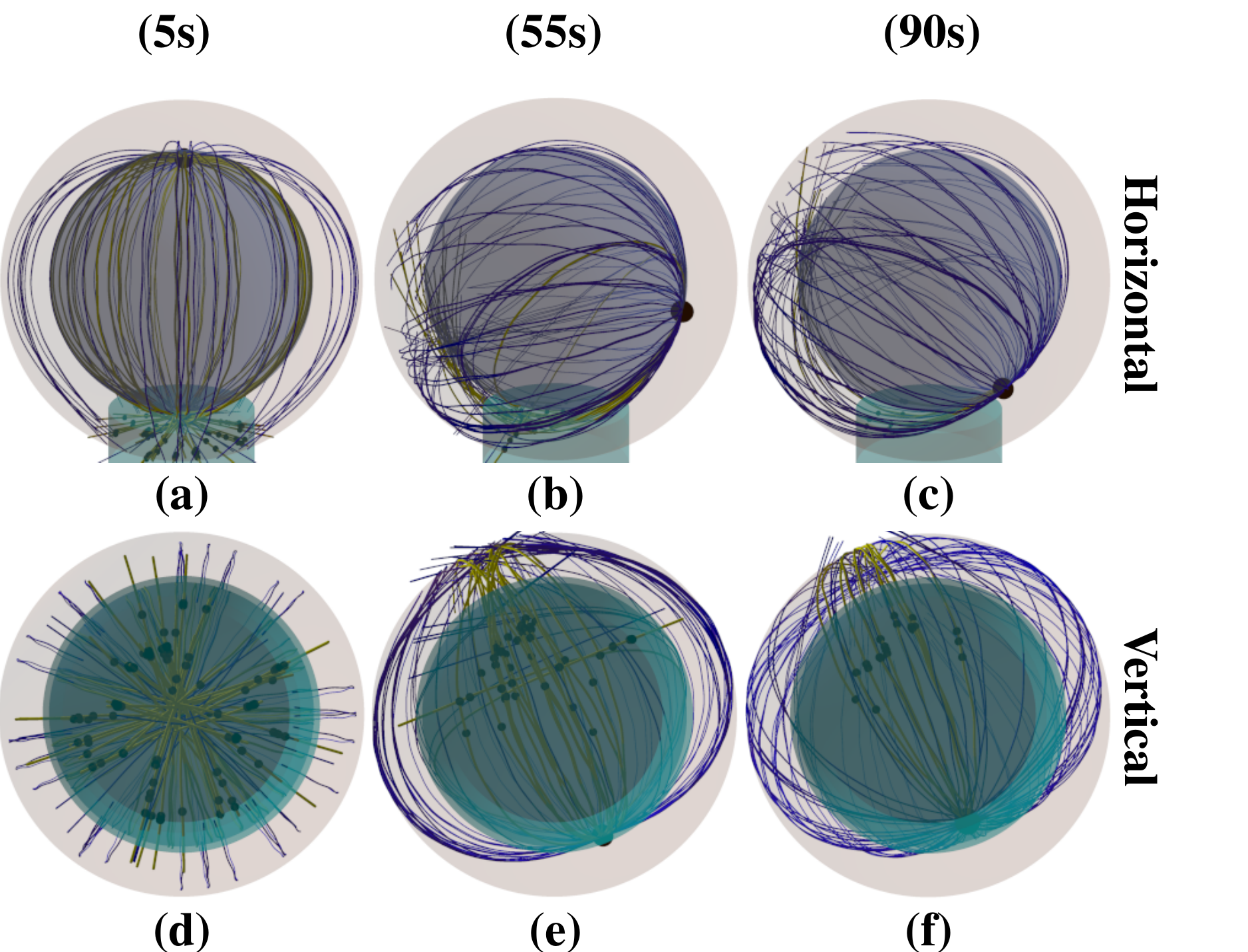}    
   \caption{Snapshots from the time-evolution of the MT cytoskeleton 
    configuration under the effect of cortical sliding mechanism with a low dynein density, 
   $\tilde{\rho}_{\textrm{IS}} = 60\mu \textrm{m}^{-2}$.
The cyan cylinder indicates the IS area. 
Blue and yellow lines are unattached and attached MTs, respectively.
The black spheres in the IS are the positions of dynein attached to MTs. 
 (a)(d)$\bar{d}_{\textrm{MIS}} = 9\mu\textrm{m}$. Originally, the attached microtubules aim from the MTOC in every direction. (b)(e)$\bar{d}_{\textrm{MIS}} = 4.5\mu\textrm{m}$. MTs attached to dynein aim predominantly in one direction. (c)(f)$\bar{d}_{\textrm{MIS}} = 1.5\mu\textrm{m}$. Just a few MTs remain under the actions of the cortical sliding and they rarely touch the surface of the cell in the IS.
   \label{fig:snapchots_cortical_region1}
}
\end{figure*}

For low, medium and high densities one observes for each a different
characteristic behavior.
In the regime of low dynein densities,
$\tilde{\rho}_{\textrm{IS}}<100\mu\textrm{m}^{-2}$,  
the repolarization velocity increases with the dynein density 
and the MTOC moves 
directly to the IS, see Fig. \ref{fig:cortical_region_1}a.
For medium dynein densities, $100\mu\textrm{m}^{-2} =<\tilde{\rho}_{\textrm{IS}}<500\mu\textrm{m}^{-2}$, the MTOC movement
is more complex, see Fig. \ref{fig:cortical_region_2}a.
For high dynein densities,
$\tilde{\rho}_{\textrm{IS}}>500\mu\textrm{m}^{-2}$,
the repolarization velocity surprisingly decreases with  $\tilde{\rho}_{\textrm{IS}}$,
see Fig. \ref{fig:cortical_region_3}a.

\subsubsection{Cortical sliding with low dynein densities}

\begin{figure*}       
     \includegraphics*[trim=40 490 50 72,clip,width=6.75in]{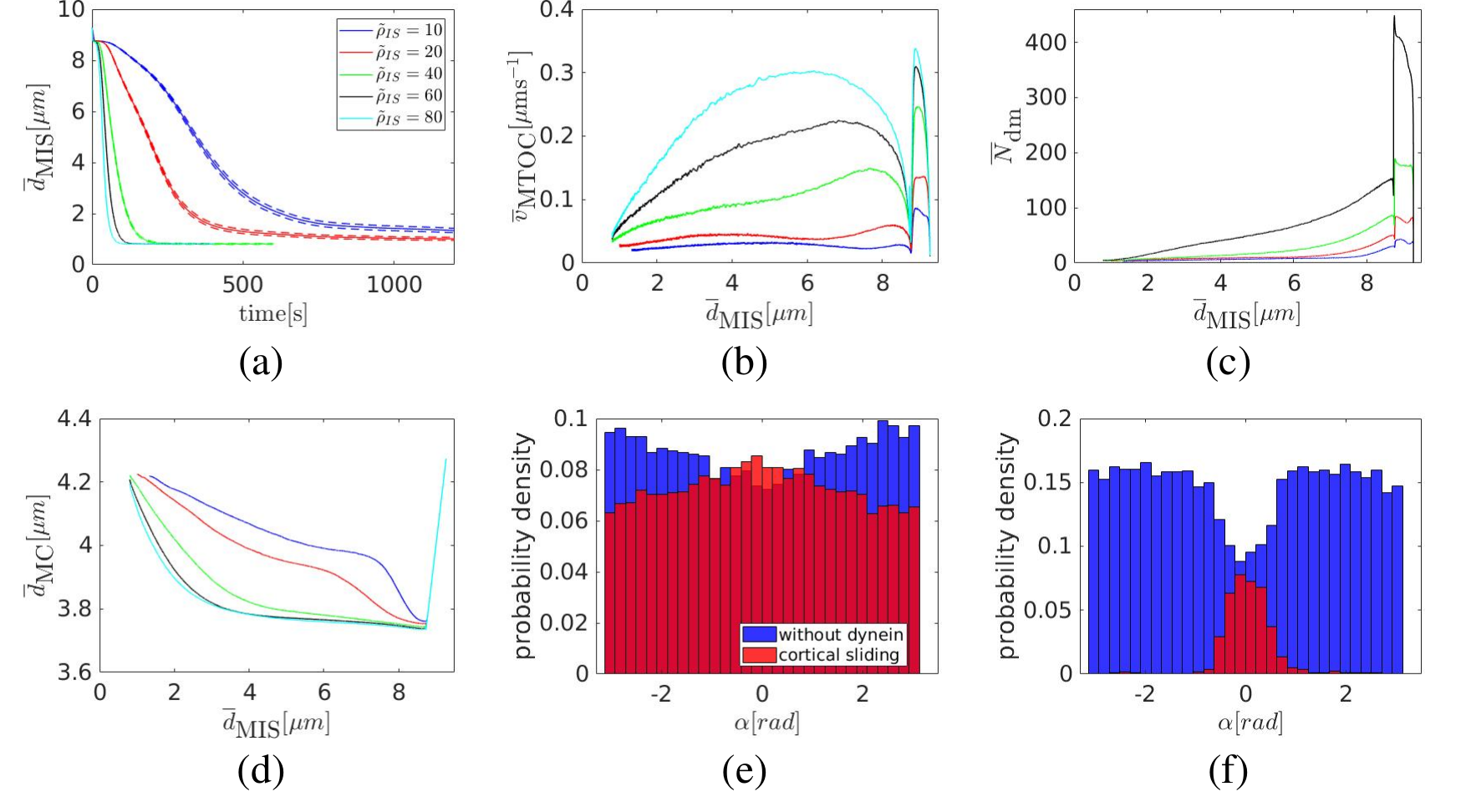}    
      \caption{Cortical sliding with low dynein densities $\tilde{\rho}_{\textrm{IS}}$. 
(a) The dependence of the average  MTOC-IS distance $\bar{d}_{\textrm{MIS}}$  on time.
(b)(c)(d) Dependencies  of the average MTOC velocity $\bar{v}_{\textrm{MTOC}}$ (b), number of dyneins acting on MTs  $\bar{N}_{\textrm{dm}}$ (c) and the
MTOC-center distance  $\bar{d}_{\textrm{MC}}$ (d) on the average MTOC-IS distance.
(e)(f)    
Probability distributions
 of the angles $\alpha$ between the first MT segment 
and the MTOC motion, $\tilde{\rho}_{\textrm{IS}} = 60\mu\textrm{m}^{-2}$. (e) $t = 5\textrm{s}$, $\bar{d}_{\textrm{MIS}} \sim 9\mu\textrm{m}$. (f) $t = 65\textrm{s}$, $\bar{d}_{\textrm{MIS}} \sim 5\mu\textrm{m}$.
\label{fig:cortical_region_1}}
 \end{figure*}

The supplementary Movie $\textrm{S3}$ shows the MTOC repositioning under the effect of the
cortical-sliding mechanism with $\tilde{\rho}_{\textrm{IS}} = 60\mu \textrm{m}^{-2}$. It shows MTs sprouting in all directions in the initial stage,   
the subsequent stalk formation and the final slowing down of the MTOC.
In Fig. \ref{fig:cortical_region_1}b the dependence of the MTOC velocity on the MTOC-IS distance is shown.
As in the case of the capture-shrinkage mechanism, the time-evolution
of the MTOC velocity can be divided into three phases. 
However, the transition points between the second and the third phase depend on the density $\tilde{\rho}_{\textrm{IS}}$. 
Similarly to the case of the capture-shrinkage mechanism, the behavior in the first phase can be explained by the interplay of fast attaching dyneins and forces from the nucleus.
In the second phase, 
the velocity of the MTOC increases in spite of a continuously decreasing number of attached dyneins, see Figs.
\ref{fig:cortical_region_1}b and \ref{fig:cortical_region_1}c,
which is due to the alignment of the MTs.
Initially, attached MTs aim in all directions, see Fig. \ref{fig:cortical_region_1}e, as for the capture-shrinkage mechanism, 
c.f. \ref{fig:snapchots_cortical_region1}a and 
\ref{fig:snapchots_cortical_region1}d.
Consequently, MTs, whose original orientation does not correspond to the movement of the MTOC, detach from dynein, see 
Figs. \ref{fig:cortical_region_1}f, \ref{fig:snapchots_cortical_region1}b and \ref{fig:snapchots_cortical_region1}e.
The probability density in the intermediate state of the repositioning
($\bar{d}_{\textrm{MIS}} \sim 5\mu\textrm{m}$) shows that attached MTs are aligned and less in numbers.
The MTOC does not significantly recede from the nucleus at the end of the repositioning, see Fig. \ref{fig:cortical_region_1}d, which implies that the 
MTs do not follow the cell membrane (with the capture-shrinkage mechanism MTs always touch the membrane), see Figs. \ref{fig:snapchots_cortical_region1}c and \ref{fig:snapchots_cortical_region1}f.
Consequently, the attachment probability is lower and leads to the decrease in
the velocity in the third phase.

\subsubsection{Cortical sliding with medium dynein densities}
 \label{Second_cortical_sliding}

\begin{figure*}
     \includegraphics*[trim=15 370 40 60,clip,width=6.75in]{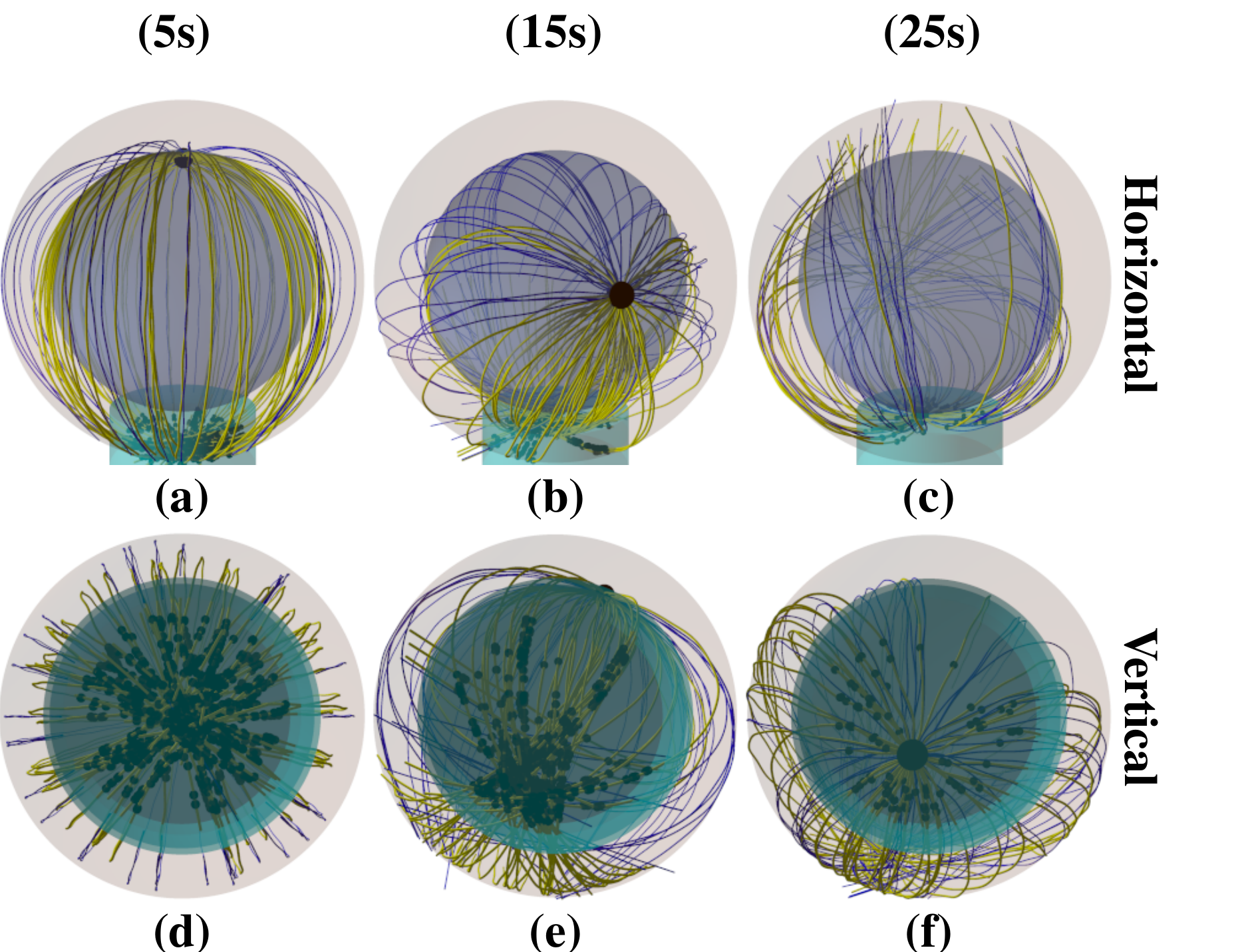}    
   \caption{Snapshots from the time-evolution of the MT cytoskeleton   
    configuration under the 
   effect of the cortical sliding alone, with a medium area density of the dynein 
   $\tilde{\rho}_{\textrm{IS}} = 200\mu \textrm{m}^{-2}$) from two perspectives.
(a)(c) $\bar{d}_{\textrm{MIS}} = 9\mu\textrm{m}$. MTs sprout from the MTOC in all directions.   
(b) (d)$\bar{d}_{\textrm{MIS}} = 5\mu\textrm{m}$. The majority of MTs is attached and the MT cytoskeleton
is deformed.
(c) (e) $\bar{d}_{\textrm{MIS}} = 1\mu\textrm{m}$. At the end of the repositioning, the MTOC passed the center of the IS and attached MTs aim in all directions.
\label{fig:snapchots_cortical_region2} 
}
\end{figure*} 

\begin{figure*}       
\includegraphics*[width=6.75in]{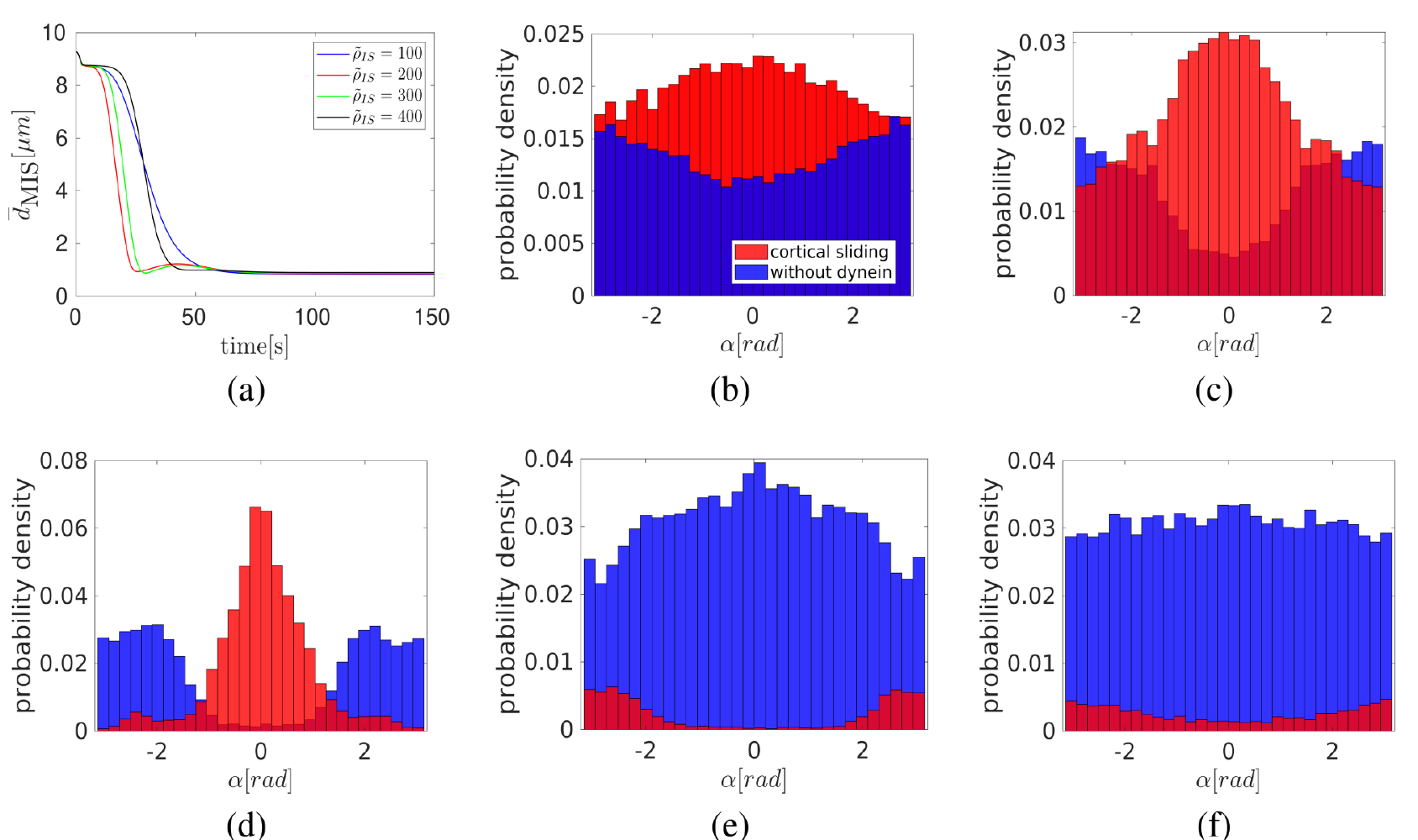}    
      \caption{Cortical sliding mechanism with medium dynein densities $\tilde{\rho}_{\textrm{IS}}$.
      (a) The dependence of the MTOC-IS distance $\bar{d}_{\textrm{MIS}}$  on time.
Probability distributions
 of the angles $\alpha$ between the first MT segment and the MTOC motion, $\tilde{\rho}_{\textrm{IS}} = 200\mu\textrm{m}^{-2}$. 
(b)$t = 5\textrm{s}$, $\bar{d}_{\textrm{MIS}} \sim 9\mu\textrm{m}$. 
(c)$t = 15\textrm{s}$, $\bar{d}_{\textrm{MIS}} \sim 6\mu\textrm{m}$.
(d)$t = 20\textrm{s}$, $\bar{d}_{\textrm{MIS}} \sim 2.5\mu\textrm{m}$.
(e)$t = 25\textrm{s}$, $\bar{d}_{\textrm{MIS}} \sim 1.5\mu\textrm{m}$, other side of IS. 
(f)$t = 60\textrm{s}$, $\bar{d}_{\textrm{MIS}} \sim 0.8\mu\textrm{m}$. 
\label{fig:cortical_region_2}}
 \end{figure*}

The differences between the behavior with low and medium dynein densities 
for the cortical sliding mechanism are analyzed in this section.
The supplementary Movie $\textrm{S4}$ shows the repositioning 
with $\tilde{\rho}_{\textrm{IS}} = 200\mu \textrm{m}^{-2}$.
The repositioning is very fast and the MT cytoskeleton
 is considerably deformed.
Moreover, the MTOC passes the IS and subsequently returns to the center of IS.
Five seconds after the initialization MTs in all directions are attached,
see Fig. \ref{fig:snapchots_cortical_region2}a and \ref{fig:snapchots_cortical_region2}d, but the direction of the MTOC motion is already established, see Fig. \ref{fig:cortical_region_2}b.
Contrary to the case of low densities, the dynein forces are sufficiently strong to hold attached MTs.
Subsequently, some MTs do not detach but take a direction partially 
aligned to the MTOC movement, see Figs. \ref{fig:cortical_region_2}c and \ref{fig:cortical_region_2}d.
Moreover, almost all MTs aligned with the MTOC motion are attached to dyneins,
compare Figs. \ref{fig:cortical_region_2}d and \ref{fig:cortical_region_1}f.
Consequently, the large majority of MTs are aligned with the direction of movement of the MTOC movement, 
causing a substantial increase of the MTOC velocity.
By comparing the temporal evolution of the MTOC-IS distance $\overline{d}_{IS(t)}$
for small, medium and large dynein densities, Figs. \ref{fig:cortical_region_1}a, \ref{fig:cortical_region_2}a and \ref{fig:cortical_region_3}a one observes 
that the velocity of the MTOC is maximal for medium densities of cortical 
sliding dyneins (see also Fig. \ref{fig:capture_cortical_comparison}a and 
\ref{fig:capture_cortical_comparison}b below).  Moreover, by comparing 
the configuration snapshots for low and high densities, Fig. \ref{fig:snapchots_cortical_region1}b,e and 
\ref{fig:snapchots_cortical_region2}b,e, one observes that the strong forces exerted at high dynein 
densities cause large deformations of the MT cytoskeleton.

Due to the deformation of the cytoskeleton, a large number of MTs is attached to dynein at the end of the repositioning, see 
Fig. \ref{fig:cortical_region_2}d and
dyneins are predominantly found at the opposite side of the IS(compared to the MTOC).
Due to the attachment, the MTOC passes the center of the IS, see Fig. \ref{fig:cortical_region_2}a and anchor points of certain dynein motors,
see Figs. \ref{fig:snapchots_cortical_region2}c and
\ref{fig:snapchots_cortical_region2}f.
The MTs are attached to anchor points; so, the probability density of $\alpha$ changes and the majority of MTs are behind the MTOC, see Fig. \ref{fig:cortical_region_2}e.
When the MTOC recurs to the IS, the probability density levels, 
see Fig. \ref{fig:cortical_region_2}f and dynein detach.

\subsubsection{Cortical sliding with high dynein densities}
\label{Third_cortical_sliding}

\begin{figure*}      
\includegraphics*[trim=40 490 50 72,clip,width=6.75in]{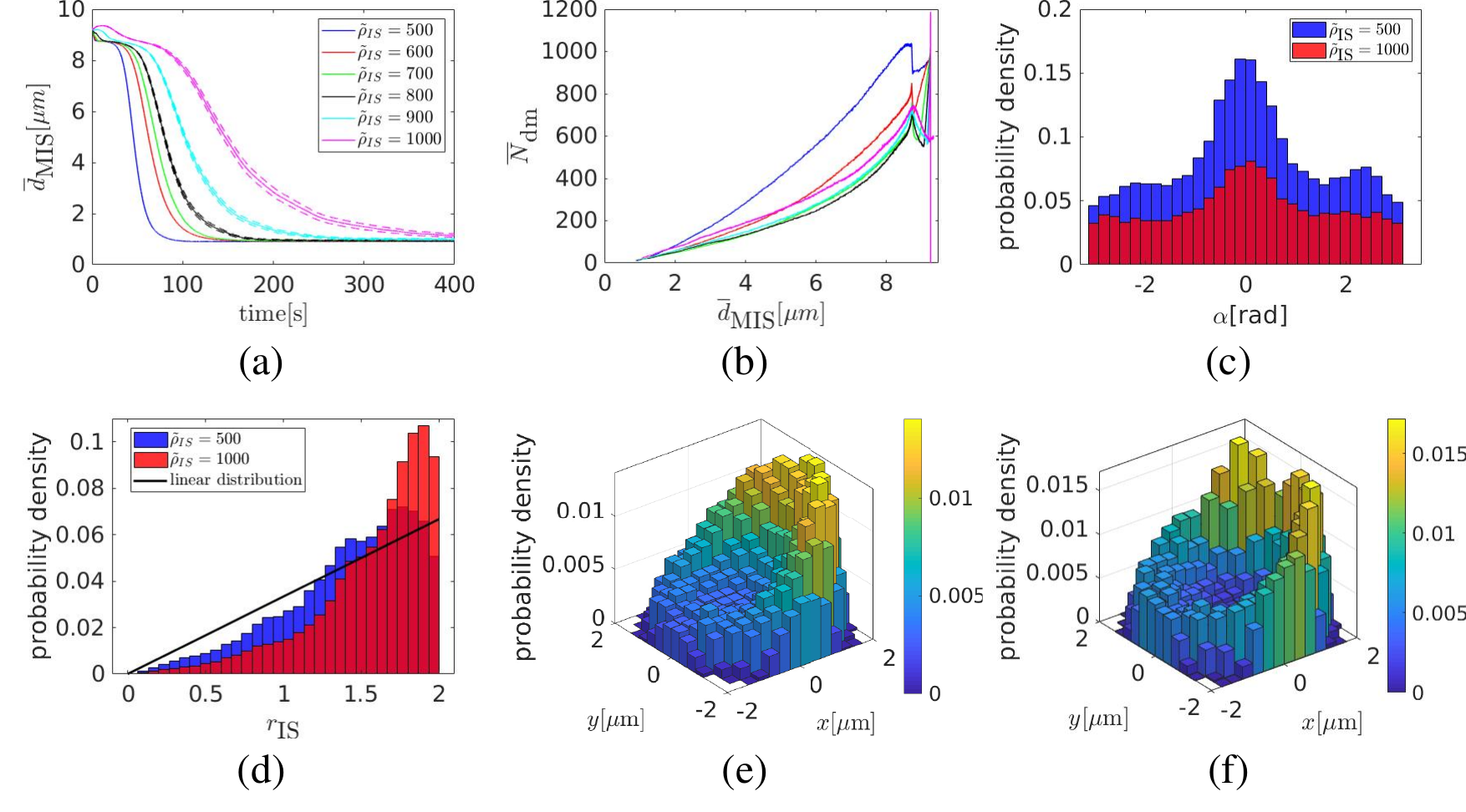}        
      \caption{Cortical sliding with high dynein densities $\tilde{\rho}_{\textrm{IS}}$.
      (a) The dependence of the average MTOC-IS distance $\bar{d}_{\textrm{MIS}}$  on time.
 (b) The dependence of the average number of dyneins  $\bar{N}_{\textrm{dm}}$ on the MTOC-IS distance. 
(c)    
Probability distribution
 of the angle $\alpha$ between the first MT segment of attached MTs and the direction of the MTOC motion, 
$\bar{d}_{\textrm{MIS}} \sim 5 \mu\textrm{m}$.
(d) Probability distribution of the distance of attached dynein anchor points from the axis of the IS $r_{\textrm{IS}}$ when $\bar{d}_{\textrm{MIS}} \sim 5\mu\textrm{m}$. 
 Two dimensional probability density of attached dynein in the IS,
$\bar{d}_{\textrm{MIS}} = 5\mu\textrm{m}$. 
(e) Area density of cortical sliding dyneins $\tilde{\rho}_{\textrm{IS}} = 500\mu\textrm{m}^{-2}$.
(f) $\tilde{\rho}_{\textrm{IS}} = 1000\mu\textrm{m}^{-2}$. 
\label{fig:cortical_region_3}}
 \end{figure*}

An example for repositioning under the effect of cortical sliding with a
high dynein density $\tilde{\rho}_{\textrm{IS}}
= 1000\mu\textrm{m}^{-2}$ is shown in the supplementary Movie $\textrm{S5}$.
As the area density $\tilde{\rho}_{\textrm{IS}}$ rises, the MTs are more and more attached at the periphery, see Fig. \ref{fig:cortical_region_3}d. 
This is further demonstrated by Figs. \ref{fig:cortical_region_3}e and 
\ref{fig:cortical_region_3}f(the center of the IS is almost depopulated
when $\tilde{\rho}_{\textrm{IS}} = 1000\mu\textrm{m}^{-2}$).
The reason is that
there is a sufficient number of dyneins to firmly catch the MT passing just the periphery of the IS.
Higher number of MTs also logically means bigger pulling forces on MTs. 
In a spherical cell, dyneins act in a competition which leads to the dynein detachment.
The bigger the competition is, more frequent is the detachment as can be seen in Fig. \ref{fig:cortical_region_3}b, where the highest number of attached dynein corresponds to the lowest area density.

Constantly attaching and detaching dynein does not allow MTs to align
with the direction of the MTOC movement.
Subsequently, the  MTOC "lingers" behind the nucleus before it moves to the IS as the dominant orientation of attached MTs forms slowly.
The duration of this inactivity rises with $\tilde{\rho}_{\textrm{IS}}$,
see Fig. \ref{fig:cortical_region_3}a.
Even when the dominant direction is established, MTs are still attached 
in every direction slowing down the movement, see Fig. \ref{fig:cortical_region_3}c.
Therefore, the slowing in the third section, c.f. Fig. \ref{fig:cortical_region_3}a, is caused by two effects.
First, the misalignment of MTs resulting in contradictory pulling forces and the lower number of attached dyneins.
Second, the increasing probability of the attachment at the periphery results
in MTs being pulled to different places contradicting
each other increasingly as the MTOC approaches the IS.

\subsection{Comparison of cortical sliding and capture-shrinkage}
\label{Comparison_1_pi}
In this section, two mechanisms are compared in terms of MTOC velocity, times and final MTOC-IS distances.
The biological motivation is that the velocity(times) indicates the efficiency of transmission of the force of dynein on the cytoskeleton and the final distance determines the completion of the repositioning.
In the previous sections, the repositioning was divided into three phases
 based on the MTOC velocity, see Figs. \ref{fig:capture_shrinkage_1}
and \ref{fig:cortical_region_1}, 
 which enabled the analysis of the dynamics based on the attached dyneins and deformations of the cytoskeleton structure.

To analyze average velocity, the repositioning is divided into three phases based on the MTOC-IS distance: the activation, the first and the second phase.
This approach will later enable a comparison with experimental results. 
The activation phase ends when $d_{\textrm{MIS}}<=8.2\mu \textrm{m}$
(identical with the first phase based on the MTOC velocity).
Although the activation phase is important for the observation of the influence of dynein motors, see Figs. \ref{fig:capture_shrinkage_1}c,
\ref{fig:cortical_region_1}c and
\ref{fig:cortical_region_3}b, 
the phase lacks experimental analogy, since in reality the IS alongside with a high dynein area density is not created instantly. 
Therefore, it will not be further analyzed. 
In the first phase, the MTOC-IS distance 
$8.2\mu \textrm{m}>d_{\textrm{MIS}} > d_{\textrm{f}} + 1\mu\textrm{m}$, where 
$d_{\textrm{f}}$ is the final MTOC-IS distance, which depends on the area density and mechanism.
The second phase comprises the last $\mu\textrm{m}$ of the MTOC journey.

The MTOC velocity in the capture-shrinkage repositioning increases with the area density of dyneins for both phases, see Fig. \ref{fig:capture_cortical_comparison}a.
The development of the average MTOC velocity of the cortical sliding repolarization is more difficult since it rises to its maximum(middle densities), see Fig. \ref{Second_cortical_sliding}
and then it falls sharply. 
The velocity of the cortical sliding repositionings is lower except when considering middle area densities of the cortical sliding dynein. 
Moreover, for the low and high densities, the velocity of the capture-shrinkage is more than two times the velocity of the cortical sliding, see Fig.
\ref{fig:capture_cortical_comparison}a.
The times of repositionings evolve accordingly, see Fig.
\ref{fig:capture_cortical_comparison}c.
 Times are longer for the case of the capture-shrinkage only when $\rho$ corresponds to the 
 middle densities of the 
 cortical sliding, see Fig.\ref{fig:capture_cortical_comparison}a.
 
 \begin{figure*}[ht]
\includegraphics*[width=6.75in]{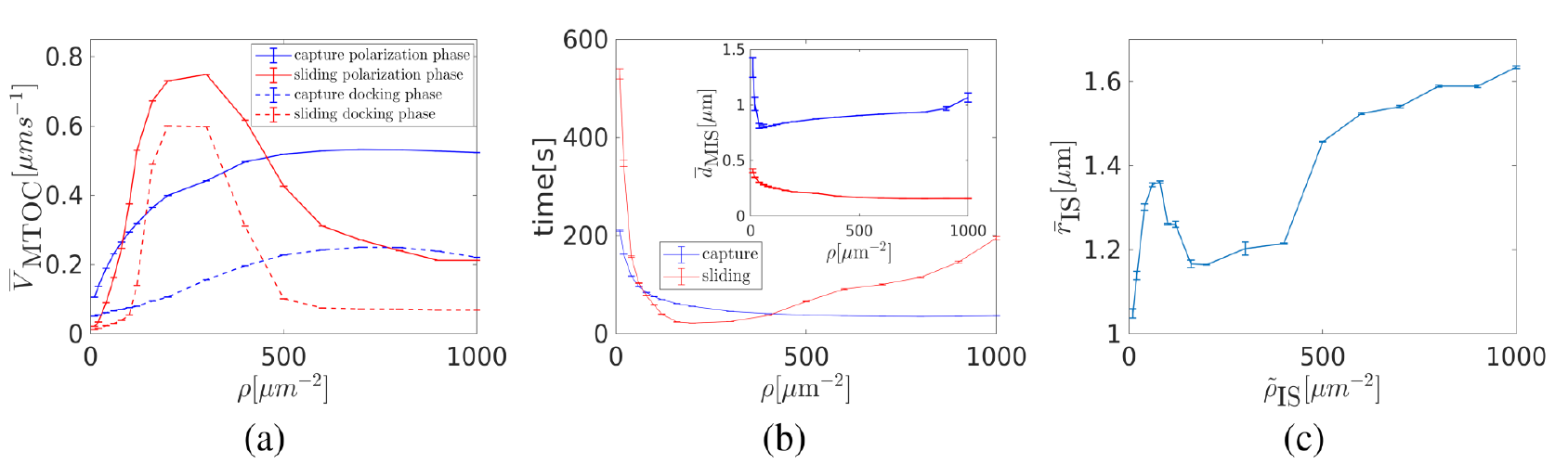}
   \caption{Comparison of the capture-shrinkage and the cortical sliding
   mechanism in terms of the average MTOC velocity in both phases $\bar{V}_{\textrm{MTOC}}$, times of repositioning and the final MTOC-IS distance 
   $\bar{d}_{\textrm{MIS}}$.
(a) The MTOC velocity in the first and the second phase. 
(b) Repositioning times. Final MTOC-IS distances in the inset.
(c) The dependence of the average distance $\bar{r}_{\textrm{IS}}$ of attached dynein motors from the axis of the IS on dynein area density $\tilde{\rho}_{\textrm{IS}}$ for the case of the sole cortical sliding.
    \label{fig:capture_cortical_comparison} } 
 \end{figure*}
 
The final MTOC-IS distance decreases with the rising $\rho$ in the case of the sole capture-shrinkage, see Fig. \ref{fig:capture_cortical_comparison}b.
In the case of the cortical sliding, the situation is more complicated due to the lack of anchor point.
The large final distances at of low area densities are caused by the insufficient pulling force.
The shortest distance is at the end of low area densities
$\rho = 80\mu\textrm{m}^{-2}$ which is caused by the fact that the formation of the narrow MT stalk, in which MTs pull in alignment,
 is limited just to low densities, see Figs. \ref{fig:cortical_region_1}e,
 \ref{fig:cortical_region_2}c and \ref{fig:cortical_region_3}c.
Then we can observe a steady rise in final distances caused by the
growing attachment of MTs at the peripheries as 
$\tilde{\rho}_{\textrm{IS}}$  ,
causes the increasing competition of pulling forces in the final stages of
the polarization.

Fig. \ref{fig:capture_cortical_comparison}c explains the lower
MTOC velocity for cortical sliding. 
First of all, let us notice that the three regimes of the cortical sliding  behavior are visible in \ref{fig:capture_cortical_comparison}c. 
We can see that the increasing $\tilde{\rho}_{\textrm{IS}}$
causes the MT attachment on the periphery of the IS, as was already suggested by 
Figs. \ref{fig:cortical_region_3}d
\ref{fig:cortical_region_3}e and \ref{fig:cortical_region_3}f.
Moreover, the attached dynein is always predominantly at the periphery, since
the average distance for the uniform distribution of dynein is $\bar{r}_{\textrm{IS}} = \frac{1}{2}R_{\textrm{IS}} = 1\mu \textrm{m}$.
Therefore, as the MTOC approaches the IS, MTs are pulled to different locations, the forces of dynein oppose each other and cause the dynein detachment.

The capture-shrinkage mechanism is faster, with the relatively narrow exception of the middle area densities. 
The cortical sliding never achieves shorter distances in comparison to the capture-shrinkage; moreover, in the case of high or low densities,
the final distances differ substantially.
Fig. \ref{fig:capture_cortical_comparison} shows the dependencies on area density.
Nevertheless, in the case of the capture-shrinkage, we consider just the density in
the center of IS. 
We should remind that radii of the center and the entire IS are $R_{\textrm{CIS}} = 0.4\mu\textrm{m}$ and $R_{\textrm{IS}} = 2\mu\textrm{m}$. 
Since the number of dyneins depends on the area, the number of dyneins in the  IS $N_{\textrm{IS}} = 25 * N_{\textrm{CIS}}$, where $N_{\textrm{CIS}}$ is the number of dynein in the center of the IS.
Consequently, Fig. \ref{fig:capture_cortical_comparison} confirms, that the capture-shrinkage mechanism is the main driving force of the repositioning since this mechanism produces bigger or comparable velocities with just $4\%$ of dynein motors of the cortical sliding.  
Moreover, the MTOC comes closer to the IS meaning that the capture-shrinkage mechanism is more likely to finish repositioning.
To summarize, considering the lower number of dynein, the capture-shrinkage
mechanism is largely superior in the considered setup. The most important difference between the two mechanisms is the firm, narrow anchor point 
in case of the capture-shrinkage mechanism.
It assures a firm attachment of the MTs, see Fig. \ref{fig:capture_shrinkage_1}f,
and a geometrical alignment of the pulling forces in all 
stages of repositioning.
The capture-shrinkage mechanism was identified as the main driving force of the repositioning \cite{yi_centrosome_2013} and
our model fully support this statement.
In the next section we will scrutinize the role of cortical sliding.

\begin{figure*}[t]
\includegraphics*[width=6.75in]{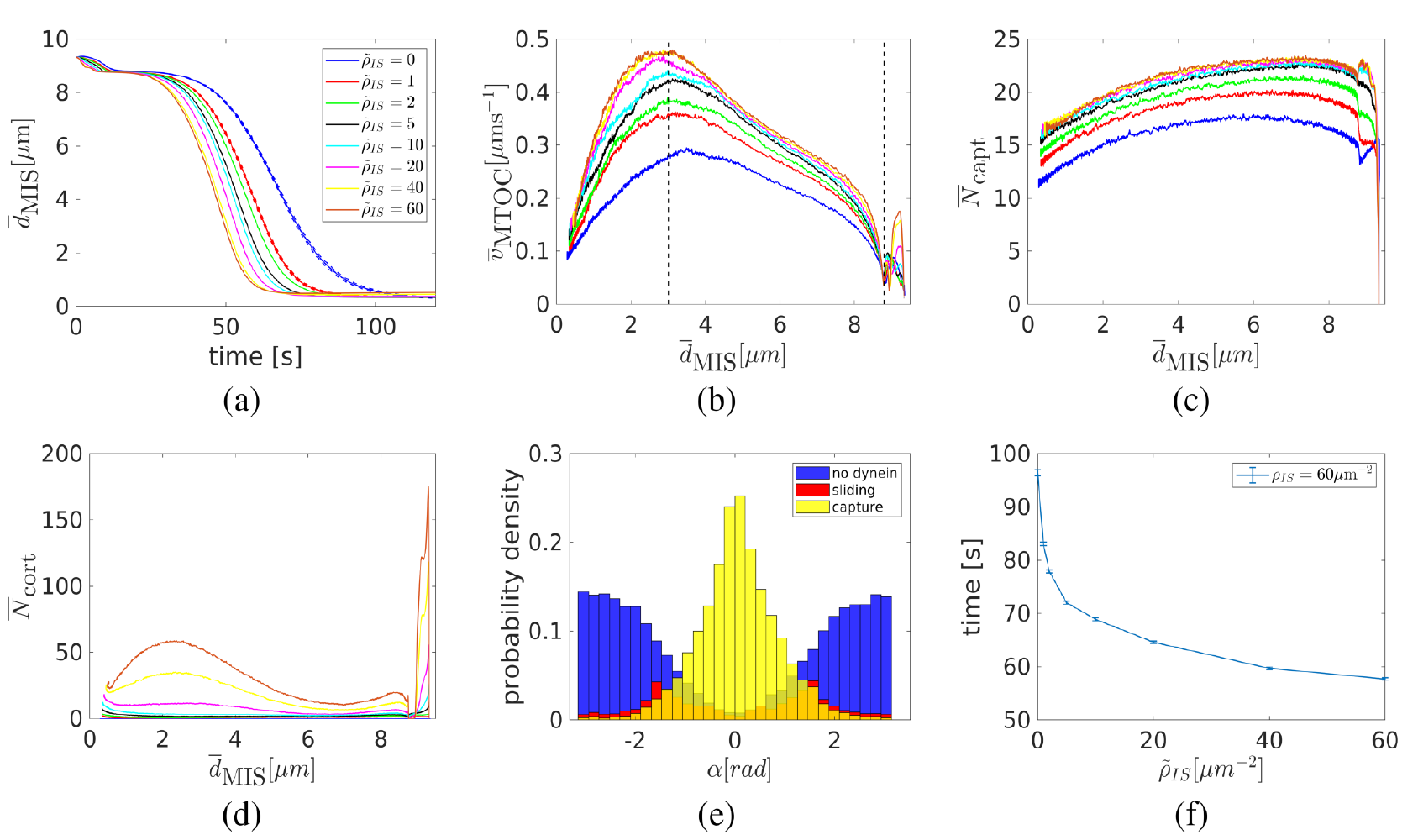}
   \caption{Combination of the capture-shrinkage and the cortical sliding:
(a) Dependence of the average MTOC-IS distance $\bar{d}_{\textrm{MIS}}$ on time.
Dependence of the average MTOC velocity $\bar{v}_{\textrm{MTOC}}$ (b), the average number of  attached capture-shrinkage dyneins $\bar{N}_{\textrm{capt}}$ (c) and the
average number of attached cortical sliding dyneins $\bar{N}_{\textrm{cort}}$ (d) on the average MTOC-IS distance (cortical sliding densities 
corresponding to different line colors in b-d are the same as in a).
(e) Probability density of the angles $\alpha$ between the first MT segment and the direction of the MTOC motion, $t = 50\textrm{s}$, $\bar{d}_{\textrm{MIS}} \sim 5\mu\textrm{m}$, $\tilde{\rho}_{\textrm{IS}} = \rho_{\textrm{IS}} = 60\mu\textrm{m}^{-2}$.
(f) Dependence of times of repositionings on cortical sliding area density $\tilde{\rho}_{\textrm{IS}}$.
       \label{fig:Capture_Cortical} } 
 \end{figure*}

\subsection{Combination of Capture-Shrinkage and Cortical Sliding}
\label{Capt_Cort_together_IS}
In this section, the interplay between two mechanisms is analyzed.
The comparison of the supplementary Movies $\textrm{S6}$ 
(capture-shrinkage, with $\rho_{\textrm{IS}} = 60\mu\textrm{m}^{-2}$,
$\tilde{\rho}_{\textrm{IS}} = 0\mu\textrm{m}^{-2}$) 
and $\textrm{S7}$ (both mechanisms combined, $\tilde{\rho}_{\textrm{IS}} = \rho_{\textrm{IS}} = 60\mu\textrm{m}^{-2}$) demonstrates the difference between the
MT cytoskeleton
 dynamics under the effect of the capture-shrinkage alone and 
under the effect of both mechanisms combined. 
The Movies show the first
few seconds
of the process.
In the case of sole capture-shrinkage, just long enough MTs attach to the center of the IS.
One clearly sees in $\textrm{S7}$ that MTs intersecting the IS and
attached to cortical sliding dynein, are passed to the center of the IS,
where they are captured by cortical sliding dynein.
The supplementary Movie $\textrm{S8}$ shows 
the complete repositioning of the MTOC under the effect of both 
mechanisms combined ($\tilde{\rho}_{\textrm{IS}} = \rho_{\textrm{IS}} = 60\mu\textrm{m}^{-2}$)

\begin{figure*}[h]
\includegraphics*[width=6.75in]{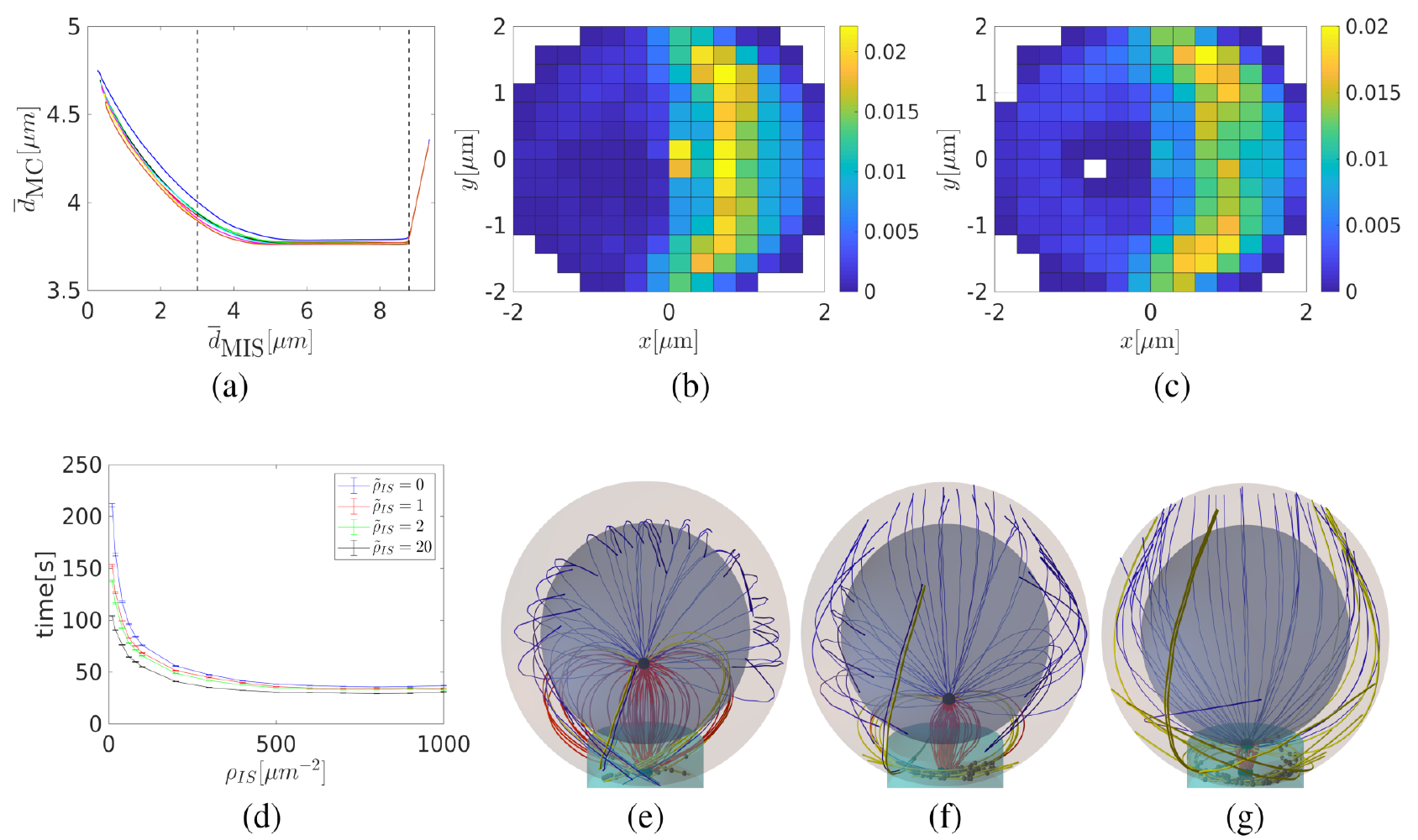}%
   \caption{
   Combination of capture-shrinkage and cortical sliding. 
   (a) Dependence of the average distance between the center of the cell and the MTOC $\bar{d}_{\textrm{MC}}$ on the average MTOC-IS distance $\bar{d}_{\textrm{MIS}}$. (b)(c) Probability density plot for the spatial distribution
   of attached dynein.
(b) $t = 50\textrm{s}$, $\bar{d}_{\textrm{MIS}} \sim 4.5\mu\textrm{m}$. (c) $t = 60\textrm{s}$, $\bar{d}_{\textrm{MIS}} \sim 1.5\mu\textrm{m}$.
(d) Repositioning times as a function of the density of
capture-shrinkage dynein $\rho_{\textrm{IS}}$ for four different values of the cortical sliding area density $\tilde{\rho}_{\textrm{IS}}$.
(e),(f),(g) Snapshots from simulation. The blue, red and bold yellow curves correspond to MTs without dynein, with the capture-shrinkage and the cortical sliding, respectively. Black dots depicts positions of attached dynein motors.
(e) $d_{\textrm{MIS}} = 4.5\mu\textrm{m}$, (f) $d_{\textrm{MIS}} = 2.5\mu\textrm{m}$,(g) $d_{\textrm{MIS}} = 1\mu\textrm{m}$
    \label{fig:Two_dimensional_probability}  }
 \end{figure*}

A quantitative analysis in Figs. \ref{fig:Capture_Cortical}a and \ref{fig:Capture_Cortical}b shows that the repolarization velocity
increases with the cortical-sliding density $\tilde{\rho}_{\textrm{IS}}$
as expected. Quite unexpectedly, it turns out that
the average number of attached capture-shrinkage dyneins depends on the 
density of cortical sliding dyneins, $\tilde{\rho}_{\textrm{IS}}$,
and increases with it, as demonstrated by
Fig. \ref{fig:Capture_Cortical}c.

This finding indicates a synergy of the two mechanisms, capture-shrinkage
and cortical sliding, and can be explained by the alignment of 
the MTs during repositioning.
The MTs attached to the cortical sliding dyneins tend to align 
with the MTOC movement as demonstrated by Fig. \ref{fig:Capture_Cortical}e, where the the dominant central peak
in direction $\alpha=0$ is caused by capture-shrinkage dyneins
and cortical sliding dyneins provide only two small peaks 
from angles towards the periphery of the IS.
As MTs align with the MTOC movement, they are 
captured by the capture-shrinkage dyneins in the 
central region of the IS and the number of cortical sliding 
dyneins drops dramatically as shown in \ref{fig:Capture_Cortical}d.

A comparison of the probability densities shown in 
\ref{fig:capture_shrinkage_1}f for the capture-shrinkage alone 
and \ref{fig:cortical_region_1}f for the cortical sliding alone
reveals the mechanism
by which the cortical sliding supports the capture-shrinkage.
Fig. \ref{fig:capture_shrinkage_1}f shows that the unattached MTs are pushed back by friction forces, which leads to the opening of the MT cytoskeleton, such 
that MTs cannot intersect the narrow center of the IS any more (visualized in Fig. \ref{fig:opening_capt_shr}f).
Attached MTs align with the MTOC movement in the case of the cortical sliding, c.f. Fig. \ref{fig:cortical_region_1}f.
Therefore, when both mechanisms are combined, MTs attached by the cortical sliding dyneins are not pushed back by friction and they align with the MTOC movement and the attachment probability of the capture-shrinkage dynein increases. 
Comparing the probability density of the angle $\alpha$ for the cortical sliding alone,
Fig. \ref{fig:cortical_region_1}f, with the one for the combined
mechanisms, Fig. \ref{fig:Capture_Cortical}e, demonstrates impressively 
that most MTs attached to cortical sliding dyneins have detached and attached to
capture-shrinkage dyneins.
 
These observations suggest an answer to the question about the role of the cortical sliding: it passes the MTs to the more efficient capture-shrinkage mechanism.
Additionally, it provides a bigger pulling force than for the cortical sliding alone due to the fact that the capture-shrinkage mechanism 
also supports the cortical sliding.
By comparison of Figs. \ref{fig:cortical_region_1}c and
\ref{fig:cortical_region_3}b with Fig. \ref{fig:Capture_Cortical}d one realizes that the dependencies of the number of cortical sliding dyneins on the MTOC-IS distance are very different. 
As the MTOC approaches the IS, the number of dyneins acting on MTs decreases in the case of the sole cortical sliding, c.f. Figs. \ref{fig:cortical_region_1}c and
\ref{fig:cortical_region_3}b, but rises for the case of combined mechanisms, c.f. Fig. \ref{fig:Capture_Cortical}d.
The reason lies in the firm anchoring of MTs in the center of the IS
and the emergence of the remarkable "arc" formations of attached dynein,
c.f. Figs. \ref{fig:Two_dimensional_probability}b and 
\ref{fig:Two_dimensional_probability}c.

The velocity of the capture-shrinkage processes explains this surprising finding.
The capture-shrinkage mechanism is more efficient since the MTs shorten due to depolymerization, align with the MTOC movement and they are pulled to the same place.
Slower stepping in the cortical sliding mechanism will result in MT lengths between the MTOC and the IS far longer than the direct distance.
Therefore, MTs have to bend, see Figs. \ref{fig:Two_dimensional_probability}e-\ref{fig:Two_dimensional_probability}g, which explains the "arc" patterns of attached dynein in the IS.
In other words, firm anchoring of capture-shrinkage pushes the cortical sliding MTs against the IS, causing further attachment.
By comparison of \ref{fig:cortical_region_1}d and \ref{fig:Two_dimensional_probability}a one observes that the MTOC approaches the IS closer in the case of combined mechanisms than in the case of the cortical sliding, which is another proof of the pulling of the MTOC towards the center of the IS. We conclude that the cortical sliding mechanism supports the dominant capture-shrinkage mechanism by "passing" the MTs, and
the capture-shrinkage mechanism supports the cortical sliding mechanism by providing the anchoring and pushing the MTs against the IS.

This synergy is also indicated by Fig. \ref{fig:Two_dimensional_probability}d,
which shows the total repositioning times as a function of the density
of capture-shrinkage dynein for various fixed values of the cortical sliding 
dynein. Although the repositioning time does not decrease further for
large values of the capture-shrinkage dynein density ($\rho_{\textrm{IS}}>600\mu m^{-2}$), it can actually be decreased further by increasing cortical-sliding dynein.
Consequently, the combination of the two mechanisms with relatively low area densities is faster than the dominant mechanism alone with maximum area density (compare cases $\rho_{\textrm{IS}} = 200\mu\textrm{m}^{-2}$ with various 
$\tilde{\rho}_{\textrm{IS}}$ with the case of $\rho_{\textrm{IS}} = 1000\mu\textrm{m}^{-2}$ in Fig. \ref{fig:Two_dimensional_probability}d). 
Further parameter variations supporting this result can be found in
the supplementary information 3.3 \ref{Additional_results}.
The effect is certainly advantageous for the cell since the cortical sliding mechanism is not as efficient as the capture-shrinkage mechanism considering a large amount of dynein required.

\section{Discussion}

We have analyzed the experiments of \cite{yi_centrosome_2013} with
a mechanistic model for the relocation of the MTOC in T cells. By 
using biologically realistic values for the model parameters, 
like the number and the stiffness of MTs, dynein pulling forces 
and detachment probabilities, and cytosol viscosity we 
can recapitulate for a wide range of dynein densities the 
experimental observations of \cite{yi_centrosome_2013}.
In particular the times scale for the completion of the 
relocation process as well as the MTOC velocities predicted 
by the model agree well with the experimental results.

Our model predicts that the cytoskeleton deforms substantially during the MTOC-repositioning process due to the combined effects of the capture-shrinkage mechanism and friction forces. The captured MTs form a narrow stalk 
between the MTOC and the IS, straightening under the tension caused by 
dynein motors acting on it and causing the 
rotation of the whole MT cytoskeleton
 towards it. Concomitantly, unattached 
MTs are pushed backwards by the emerging viscous drag "opening" the 
MT cytoskeleton
, c.f. \ref{fig:opening_capt_shr}e and \ref{fig:opening_capt_shr}f.
Thus our model provides a mechanistic explanation of the MT cytoskeleton
opening that is clearly visible also in 
the the experiments, as for instance in Fig. 5A of
\citep{yi_centrosome_2013}.
The opening can also be seen in the case of combined mechanisms, although for partially different reasons \ref{fig:Two_dimensional_probability}e-\ref{fig:Two_dimensional_probability}g.

The MT cytoskeleton
 opening might have interesting consequences for the 
distribution of $\textrm{Ca}^{2+}$ in the cell, which is highly relevant for the cell function. As the cytoskeleton rotates, the mitochondria are dragged with it \citep{maccari_cytoskeleton_2016}, until they approach the IS.
Due to the MT cytoskeleton
 opening the mitochondria are positioned asymmetrically 
around the IS resulting in an asymmetric absorption and redistribution
of $\textrm{Ca}^{2+}$ by the mitochondria. Consequently, an
asymmetric distribution of $\textrm{Ca}^{2+}$ arises around the IS,
whose function might deserve further investigation.

The detailed analysis of the MT cytoskeleton
 arrangement for the
cortical sliding mechanism revealed three 
different deformation characteristics depending on three 
regimes for the dynein density \ref{Cortical_Sliding_mechanism_in_IS}.
This observation opens the interesting experimental perspective to 
estimate the dynein distribution from the MT cytoskeleton
 deformation 
during the MTOC repositioning.

Moreover, our model also predicts a biphasic behavior of the 
relocation process
as reported for the experiments in \cite{yi_centrosome_2013}.
The Figs. \ref{fig:capture_shrinkage_1}a and \ref{fig:Capture_Cortical}a
bear a clear resemblance to Figure 3D of \cite{yi_centrosome_2013}.
We showed that after a short initial period, in which MTs start to 
attach to the dynein, the first phase observed experimentally 
corresponds in our model to the circular motion  of the 
MTOC around the nucleus and the second phase to  the last $1\mu\textrm{m}$ 
of the MTOC movement, where it detaches from the nucleus and 
moves more or less straight to the IS with a substantially 
reduced velocity -- for both mechanisms, capture-shrinkage and cortical
sliding, and a large range of dynein densities.
During the latter phase the MTOC increases its distance
from the cell center by approximately $1\mu\textrm{m}\sim 
0.2 * R_{\textrm{Cell}}$, which is close to the value
reported in \cite{yi_centrosome_2013}.

It was hypothesized in \citep{yi_centrosome_2013} that a resistive force
emerges at the transition point between the phases, causing slowing down of the MTOC. Our model shows that the assumption of a resistive force is not
necessary to explain the biphasic behavior: 
the velocity of the MTOC is only determined by the number of motors 
pulling on the MTs, and on MT alignment \ref{fig:capture_shrinkage_1},
\ref{fig:cortical_region_1}a and \ref{fig:cortical_region_3}.
The reason for the slowing down is therefore simply the decrease of 
the number of dyneins attached to the MTs, which again is a consequence of 
the changing geometry and forces during the movement of the MTOC,
i.e. a consequence of the interplay between the MT cytoskeleton
 and motors.

Experimentally, it was also observed that a treatment with taxol substantially
reduced the velocity of the repositioning. Taxol impedes depolymerization
of the MTs which we could mimic in our model by reducing the 
capture-shrinkage efficiency. With this modification our model reproduces
the experimental observation (Figs. S\ref{fig:cortical_with_small_capture}a and S\ref{fig:cortical_with_small_capture}b).

An interesting prediction of our model is that the two mechanisms, 
the capture-shrinkage and the cortical sliding, appear to act in remarkable synergy \ref{Capt_Cort_together_IS},
which provides an answer to the question about the role of the
cortical sliding  \cite{yi_centrosome_2013}.
The cortical sliding passes the MTs to the more efficient capture-shrinkage 
mechanism which in return provides the firm anchor point.
Therefore, the cortical sliding is useful even in the configuration, when the capture-shrinkage is dominant.
The synergy has a very practical effect since the combination of mechanisms with relatively low area densities can be faster than only the dominant mechanism with much higher area density 
\ref{fig:Two_dimensional_probability}d.
Therefore, the synergy of two mechanisms can substantially reduce the area densities necessary for an effective repositioning and reduces the necessary
resources (dynein). 
In our model the cytoskeleton does not have to force its way through multiple organelles with complex structure and the synergy manifests itself mainly in the velocity of the repositioning process. But one could speculate
that in the real cell the synergy can actually make the 
difference between completed and no repositioning.

It was proven in \citep{combs_recruitment_2006} that dynein colocalize with the ADAP ring in the pSMAC. 
Moreover, in \cite{kuhn_dynamic_2002} was hypothesized that the MTs are 
part of the reason why the pSMAC takes the form of the ring.
Additionally, \cite{kuhn_dynamic_2002} reported the sharp turns in MTs upon the interaction with the pSMAC and that the MTs do not project directly to the cSMAC.
In our model the cortical sliding dynein is homogeneously distributed 
over the entire IS, nevertheless we observe that dyneins attach to MTs predominantly at the periphery of the IS, c.f. \ref{fig:cortical_region_3}d, \ref{fig:cortical_region_3}e and \ref{fig:cortical_region_3}f.
If both mechanisms are present attached cortical sliding dyneins 
are even completely absent in the central region, c.f.  \ref{fig:Two_dimensional_probability}.
We therefore conclude like \cite{kuhn_dynamic_2002} that cortical 
sliding MTs do not project directly into the cSMAC and identify the 
periphery of the IS as the region where cortical sliding MTs are anchored.
In agreement with the experiments \cite{kuhn_dynamic_2002} we observe that
cortical sliding MTs turn upon contact with the periphery \ref{fig:Two_dimensional_probability}e-\ref{fig:Two_dimensional_probability}g, twist and contribute to the formations of dynein "arcs".
Since the dynein in the central region of the IS does not 
contribute to the MTOC repositioning via cortical sliding, 
one could hypothesize that the pSMAC takes the shape of the 
ring to facilitate interaction with MTs \cite{kuhn_dynamic_2002}.

We presented a numerical analysis of the repositioning in the case when the MTOC and the IS are initially at the opposite sides of the cell.
Even the case so restricted brought interesting results
enabling the comparison with experiments and proposition of explanation of
experimental observable. 
We found that the cell performs the repositioning with great efficiency. 
The dyneins are placed only at the peripheries of the IS(pSMAC), which is the place where they are used the most, evacuating less used regions.
Moreover, we introduced the synergy of two mechanisms that minimizes the necessary area density of dynein.

In this work we presented the results of our theoretical analysis
of the MTOC repositioning that is relevant for the experimental 
setup in \citep{yi_centrosome_2013}, where the IS and the initial position
of the MTOC are diametrically opposed. Here it turned out that even
if both mechanism, capture-shrinkage and cortical sliding, are at work, 
capture-shrinkage is always dominant as reported in 
\citep{yi_centrosome_2013}. In a second part of this work
\cite{Hornak-Rieger-2}, we will examine other initial positions of the
MTOC and the IS that will naturally occur
in biologically relevant situations, and we will investigate under
which circumstances cortical sliding will become the dominant 
mechanism over capture-shrinkage. 
Moreover, we will further demonstrate the synergy of two mechanisms introduced in this work and prove that it has more far-reaching effects in other
initial configurations than the one studied here. Also in the situation 
in which the T cell establishes two IS interesting dynamical 
behavior of the MTOC can be expected and will be analyzed in detail.
In the end, we will see that in T cells the two mechanisms,
capture-shrinkage and cortical sliding, and the spatial distribution 
of dynein are combined such as to minimize the number of dyneins 
necessary for polarization and to minimize the damage of the MT cytoskeleton.

\section*{AUTHOR CONTRIBUTIONS}
H.R. designed the research. I.H. performed calculations, prepared figures,
and analyzed the data. I.H. and H.R. wrote the manuscript.

\section*{ACKNOWLEDGMENTS}
The authors wish to thank Bin Qu and Renping Zhao for helpful discussions.
This work was financially supported by the German Research Foundation
(DFG) within the Collaborative Research Center SFB 1027.

\end{document}


\setcounter{page}{1} 

\begin{center}
{\fontsize{17}{12}\selectfont
Stochastic model of T Cell repolarization during target elimination (I)\\
\vspace{1mm} SUPPLEMENTARY MATERIAL
}\\
\vspace{5mm}

{\fontsize{12}{12}\selectfont
Ivan Hornak, Heiko Rieger\\}
\end{center}

\section{\large{\bf{Model of the cell}}}
\label{Model_of_the_cell}

\subsection{ Microtubules }

The microtubules(MTs) are represented as semiflexible filaments, therefore
the Hamiltonian of a single MT is given by:
\begin{equation}
\label{basic_hamiltonian}
H = \frac{ \kappa }{ 2 }\int_{0}^{L}\bigg|\frac{\partial \vec{t}}{\partial s}\bigg|^{2},
\end{equation}
where $\kappa$ is bending rigidity( $2.2 *10^{-23} \textrm{N} \textrm{m}^{2}$ ), $L$ is the length of MT, $s$ is arc length, 
$\vec{t} = \frac{\partial \vec{r}}{\partial s}$ is unit tangent vector and $\vec{r}(s)$ is a position \cite{broedersz_modeling_2014}.
A single MT is represented as a chain of $N$ beads with coordinates $\vec{r}_{1}, ... , \vec{r}_{N}$ connected by $N - 1$ tangents
$\vec{ t }{ i } = \vec{ r }_{ i + 1 } - \vec{r}_{ i }$ of the length $k = L / ( N - 1 )$.
In the present model, the length $k = 0.8 \mu \textrm{m}$ was used.
Since the MT is an inextensible polymer discretized into $N$ beads,
$N-1$ constraints must be fulfilled:
\begin{align}
C^{\textrm{micro}}_{i}= | \vec{r}_{i} - \vec{r}_{i + 1}| = k && i = 1, ... , N - 1 
\end{align}

The length of the MTs varies considerably, but
since short MTs are not relevant for repositioning, we include only just MTs that reach from the MTOC to the IS in the first seconds of repositioning. 
The maximum length of a MT should be $L>\pi * R_{\textrm{Cell}}$ to always reach from MTOC to IS.
In order to reach the IS in the first stages of repolarization, the length must be  $L>\frac{3}{4} * \pi * R_{\textrm{Cell}}$.
Thus, the number of beads $N$ is uniformly distributed between 15 and 20.


\subsubsection{Bending forces of the microtubule}
\begin{figure}[ht]
  \includegraphics[trim=80 620 50 68,clip,width=\linewidth]{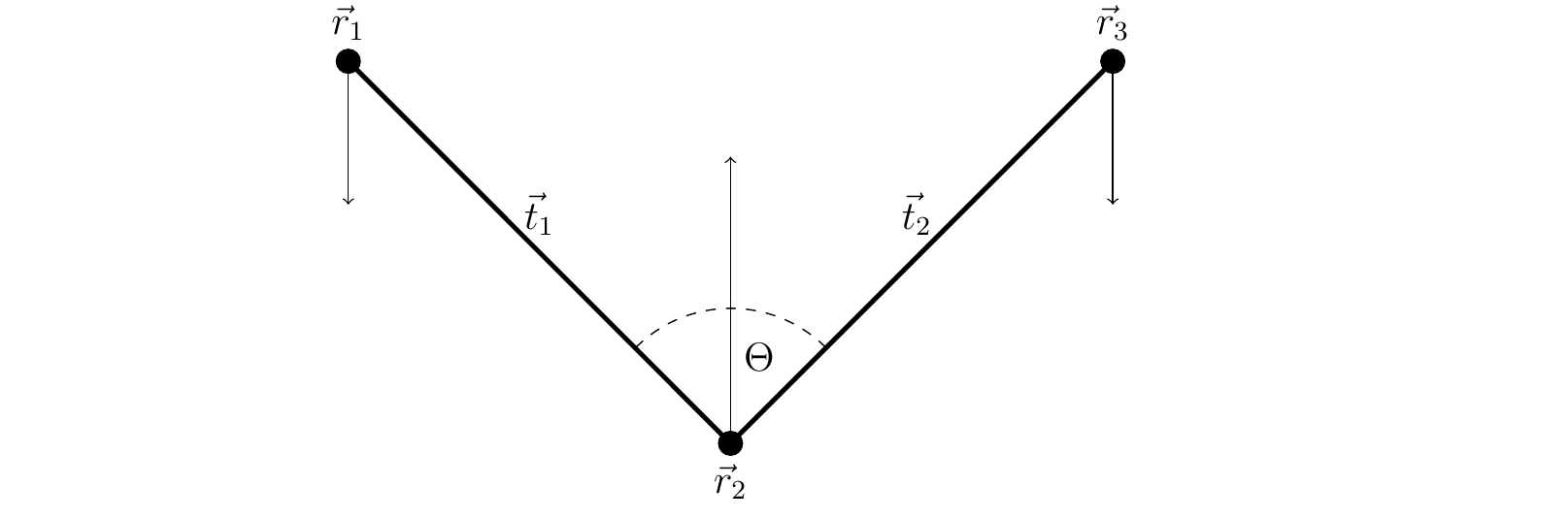}
\caption{The sketch of bending forces acting on the MT composed of three beads.
The filled circles represent the beads of the MT. 
The bending force is determined by the angle $\Theta$ between the tangents $\vec{t}_{1}$ and $\vec{t}_{2}$.
The narrow arows depict bending forces acting on beads.
 \label{bending_sketch} }
\end{figure}

The Hamiltonian for the discretized MT can be expressed as:
\begin{equation}
\label{simple_bending}
 H_{bend} = 
 \kappa_{d} \sum_{i = 0}^{ N - 2 } \left( 1 - \frac{\vec{t}_{i} \vec{t}_{i + 1} }{|\vec{t}_{i}| |\vec{t}_{i + 1}| }\right),
\end{equation}
where $\kappa_{d} = \kappa / k $ is the bending rigidity of the discretized model.
The bending force acting on  bead $i$ is the derivative of the discretized Hamiltonian with the respect to $\vec{r}_{i}$:
\begin{equation}
\vec{F}^{\textrm{bend}}_{i} = - \frac{ \partial H_{bend}}{ \partial \vec{r}_{i} }.
\end{equation}
If we consider the simplest case (sketched in Fig. \ref{bending_sketch}) of three points with coordinates $\vec{r}_{1}$, $\vec{r}_{2}$ and $\vec{r}_{3}$ connected
with the tangents $\vec{t}_{1}$ and $\vec{t}_{2}$, the bending forces acting  on the beads are:

\begin{figure}[h]
  \includegraphics[trim=80 640 50 68,clip,width=\linewidth]{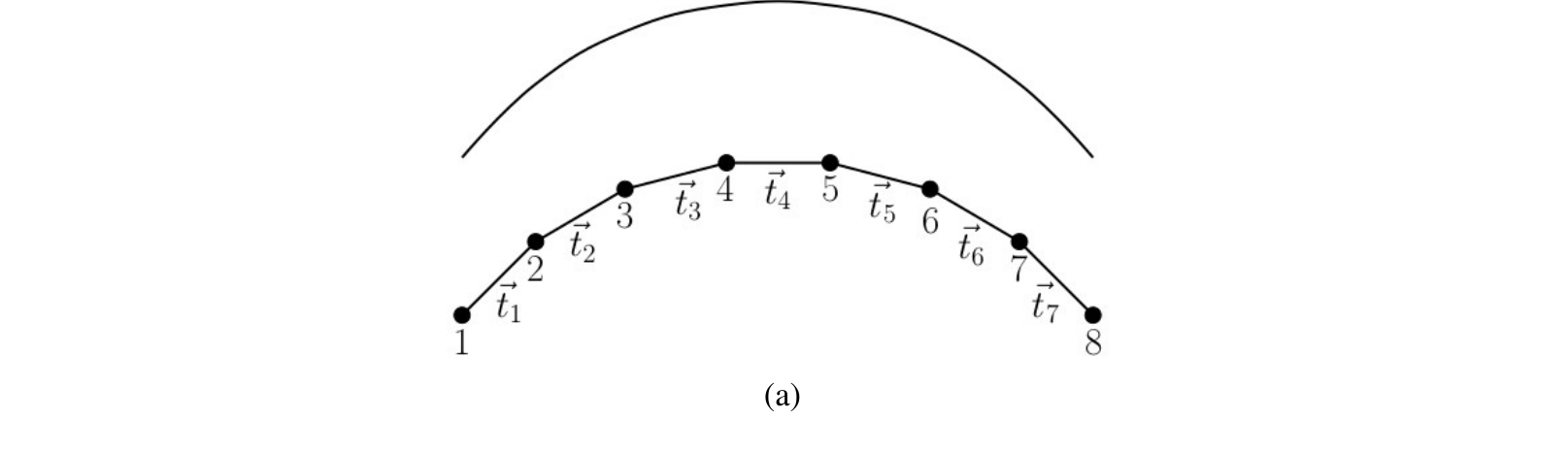}
\caption{Bead-rod model of the microtubule. 
The MT is divided into $8$ point connected by $7$ rods. 
The circles correspond to the positions of the beads. 
The lengths of rods connecting beads remain constant.  
}    
\label{micro_discretization}
\end{figure}

\begin{subequations}
\label{bending_forces}
\begin{align}
         \vec{F}^{\textrm{bend}}_{1} = \frac{ \kappa_{d} }{ |\vec{t}_{1}| }\left( - \frac{\vec{t}_{2}}{|\vec{t}_{2}|}
 + \frac{\vec{t}_{1}}{|\vec{t}_{1}|} \left( \frac{\vec{t}_{1} \cdotp \vec{t}_{2} }{ |\vec{t}_{2}| |\vec{t}_{1}| } \right)  \right),\\
         \vec{F}^{\textrm{bend}}_{3} = \frac{ \kappa_{d} }{ |\vec{t}_{2}| }\left(  \frac{\vec{t}_{1}}{|\vec{t}_{1}|}
 - \frac{\vec{t}_{2}}{|\vec{t}_{2}|} \left( \frac{\vec{t}_{1} \cdotp \vec{t}_{2} }{ |\vec{t}_{2}| |\vec{t}_{1}| } \right)  \right)\\
  \vec{F}^{\textrm{bend}}_{2} = -\vec{F}^{\textrm{bend}}_{3}  - \vec{F}^{\textrm{bend}}_{1}. 
\end{align}
\end{subequations}

  \subsubsection{Drag coefficient}

The drag force on an object depends on the speed of the object, the viscosity $\mu$ of the fluid, the size, and the shape of the object.
The intracellular viscosity is very hard to measure or estimate, since
the space between the nucleus and cell membrane is far from homogenous.
The viscosity of aqueous domains of cell cytoplasm corresponds to the viscosity of water \cite{puchkov_intracellular_2013-1}. 
On the other hand, in other microdomains, the local viscosity differs from the viscosity of water by more than three orders of
magnitude \cite{bausch_measurement_1999-1}.
The effective viscosity in our model was chosen as:
\begin{equation}
\mu \sim 30 \mu{\textrm{w}} = 0.03 \textrm{Pa} \cdot \textrm{s}, 
\end{equation}
where $\mu_{\textrm{w}} = 0.9775 \textrm{mPa} \cdot \textrm{s}$ is the viscosity 
of water at temperature of $T = 21^\circ \textrm{C}$.

The MT is divided into segments of a cylindrical shape whose length is substantially bigger than the diameter. 
The drag depends on the object's area projected normal to the direction of the motion.
Therefore, drag forces for the motions parallel and perpendicular to the object are defined as  $F = \gamma_{\parallel} \dot{x}$ and $F = \gamma_{\perp} \dot{x}$, respectively, where $\dot{x}$ is speed of the object.
Parallel $\gamma_{\parallel}$ and perpendicular $\gamma_{\perp}$ 
drag coefficients \cite{howard_mechanics_2001}  are defined for the case of a cylinder in the fluid with viscosity $\mu$ as:

\begin{subequations}
\label{drag_coefficients_basic_1}
\begin{align}
         \gamma_{\parallel} = \frac{ 2\pi \mu k }{\textrm{ln}( k / d) - 0.2}\\
	 \gamma_{\perp} = \frac{ 4\pi \mu k }{\textrm{ln}( k / d) + 0.84},
\end{align}
\end{subequations}

\noindent
where $k$ and $d$ are the length and the diameter of the cylinder, respectively. 
Therefore, the friction of the cylinder is anisotropic $\gamma_{\perp} \sim 2  \gamma_{\parallel}$.
However, the anisotropy is hard to implement, as the orientation of segments  varies.
Therefore, the anisotropy is not implemented. 
For the case of simplicity, the perpendicular drag coefficient is considered in the simulation to the beads of MT and the same drag coefficient is attributed to all MTOC points.

\subsection{Drag force of organelles}
EM, Golgi and mitochondria are the organelles with different structures and sizes.
However, their drag force can be estimated from their volume and surfaces.
A dynamic shape factor\cite{leith_drag_1987}, $K_{\textrm{sf}}$, 
can be defined to calculate the drag coefficient of the nonspherical particle:
\begin{equation}
\gamma =   3 \pi d_{v} \mu K_{\textrm{sf}},
\end{equation}
where $d_{v}$ is the diameter of the sphere with the same volume as the object.
The drag force can be divided between the form drag, coming from the pressure on the surface, and tangential shear stress.
Form drag is determined by the objects area projected normal to the direction of the motion. 
It can be expressed through the Stokes law form drag on a sphere, whose projected area equals to the projected area of a nonspherical object.
The diameter of such a sphere is $d_{n}$.
The friction force on the surface can be expressed by the friction on the sphere with the same effective surface, which has the diameter $d_{s}$.
The dynamic shape factor can be defined as:
\begin{equation}
\label{dynamic_shape}
 K_{\textrm{sf}} = \frac{d_{n} + 2 d_{s}}{3 d_{v}}. 
\end{equation}

\noindent
The number of mitochondria  were measured(just the case of one cell) in \cite{maccari_cytoskeleton_2016} (44 mitochondria in a T-Cell).
The size and shape of MTs varies greatly, since they can shrink, grow, go through fission and fusion \cite{bereiter-hahn_dynamics_1994,jakobs_high_2006,chaudhuri_cell_2016,noauthor_atlas_1966,jakobs_super-resolution_2014}.
We consider the most common spherocylindrical shape approximated by the cylinder,
whose 
diameter and the length were estimated as $0.75\mu \textrm{m}$ and $1.5\mu \textrm{m}$, respectively.
Golgi apparatus is a very complex structure composed from multiple classes of cisternae differing in form, function and composition that are stacked in various ways \cite{xu_asymmetrical_2013,ladinsky_golgi_1999,day_three-stage_2013,huang_golgi_2017}.
The endoplasmic reticulum (ER) is also a complex organelle composed from a bilayer forming nuclear envelope and a network of sheets and dynamic tubules \cite{westrate_form_2015,english_endoplasmic_2013,english_peripheral_2009,shibata_rough_2006,
hu_weaving_2011}.
Golgi, ER and mitochondria are connected to cytoskeleton \cite{gurel_connecting_2014} and \cite{maccari_cytoskeleton_2016}.

For the rough approximate evaluation of drag force we consider the estimates from \cite{alberts_molecular_2007}, table 12.
The major organelles including EM and Golgi are close to the center and they are in contact.
Consequently, effective viscosity in the regions close to the nucleus of the cell is different from more aqueous 
domain close to the cell membrane where the rotation of the cytoskeleton takes place.
Therefore, we assume that the effective viscosity of the medium in which EM and Golgi travel is $\mu_{2} = 10 * \mu$. We will express the drag coefficient of organelles as a function of viscosity and compare it with the drag coefficient of the cytoskeleton.
For the estimate of the cytoskeleton drag coefficient, the cytoskeleton of 100 MTs was considered and the 
drag coefficient were calculated using (\ref{dynamic_shape}).
\begin{align}
\label{drag_coefficients}
\gamma_{\textrm{GSER}} \sim 0.00131 \mu \\ 
\gamma_{\textrm{RER}} \sim 0.00160 \mu\\
\gamma_{\textrm{Mito}} \sim 0.00257 \mu\\
\gamma_{\textrm{Cyto}} \sim 0.00270 \mu
\end{align}

From the equations (\ref{drag_coefficients}) it can be seen that the 
$\gamma_{\textrm{total}} = ( \gamma_{\textrm{Cyto}} + \gamma_{\textrm{Mito}} + \gamma_{\textrm{RER}} + \gamma_{\textrm{GSER}}) \sim 3 * \gamma_{\textrm{Cyto}}$.
Consequently, to consider the drag force from the organelles in the cell, the drag coefficient of MT is tripled.
Thus, the equation (\ref{drag_coefficients_basic_1}) can be rewritten as:
\begin{subequations}
\label{drag_coefficients_basic}
\begin{align}
         \gamma_{\parallel} = 3 * \frac{ 2\pi \mu k }{\textrm{ln}( k / d) - 0.2}\\
	 \gamma_{\perp} = 3 * \frac{ 4\pi \mu k }{\textrm{ln}( k / d) + 0.84}.
\end{align}
\end{subequations}

\subsection{ Confinement of the cytoskeleton }
The cytoskeleton moves between the wall of the cell and the nucleus.
They have a spherical shape and they are modeled as force fields.
The force of the wall is null if $|\vec{r}_{i}|<=R$. 
Otherwise, the force acting on the bead of the MT or the MTOC is expressed as:
\begin{equation}
 \vec{F}^{\textrm{wall}}_{i} = -1 \frac{ \vec{r}_{i}}{|\vec{r}_{i}|}  k_{1}  \textrm{exp}( k_{2} (|\vec{r}_{i}| - R)), 
\end{equation}
where $R$ is the radius of the cell and  $k_{1} = 20\textrm{pN} \mu \textrm{m}^{-1}$ and $k_{2} = 1 \textrm{m}^{-1}$ are chosen constants.
The force of nucleus is null if $|\vec{r}_{i}|>R_{nucleus}$(radius of the nucleus). Otherwise, it can be expressed by:
\begin{equation}
 \vec{F}^{\textrm{nucleus}}_{i} =   \frac{ \vec{r}_{i}}{|\vec{r}_{i}|}  k_{1}  \textrm{exp}( k_{2} ( R_{nuc} - |\vec{r}_{i}|)), 
\end{equation}
where $R_{nuc}$ is the radius of the nucleus.

\subsection{Dynein motors}
\label{Dynein_motors_ref}
Unfortunately, since
the results from the measurements differ greatly, the mechanical properties of dynein remain uncertain. 
Therefore, the parameters in this section are estimations.
The dynein has an anchor and attachment points connected by a stalk.
The anchor point has a stable position and the attachment point walks on the MT. 
The force acting on the MT is determined by the length of the stalk,
whose relaxed length was estimates as 
$L_{0} = 18n \textrm{m}$
\cite{goodenough_high-pressure_1987,gee_extended_1997,goodenough_structural_1984,schmidt_insights_2012}.

Unattached dynein is represented just with one point on the surface of the cell.
If the dynein is closer to the MT than $L_{0}$,
the motor protein can attach to the filament.
Fluctuations of the membrane can move the dynein motor to the MT.
Therefore, the attachment probability is defined as:
\begin{align}
 p_{a} = 5\textrm{s}^{-1} && d_{md}<=L_{0}\\
  p_{a} = 5 \cdot \textrm{exp}( - ( d_{md}-L_{0} ) / p_{d} ) \textrm{s}^{-1} && d_{md}>L_{0},
\end{align}
where $d_{md}$ is the distance of the dynein point to the closest point of the MT 
and $p_{d} = 10^{-7}$ is a chosen parameter. 
If the MT is attached, the anchor and attachment points of the dynein motor are
placed to the same point on the MT.
Attachment probability of dynein is unknown; therefore, the attachment probability $p_{a}$ corresponding to the attachment ratio of kinesin is considered  \cite{leduc_cooperative_2004}.  
The force of dynein motor comes from the elastic properties of the stalk:
\begin{align}
 |F^{\textrm{Dynein}}_{i}| = 0, &&  |\vec{r}_{\textrm{Dynein}}| < L_{0} \\
 \vec{F}^{\textrm{Dynein}}_{i} = k_{\textrm{Dynein}} ( |\vec{r}_{\textrm{Dynein}}| - L_{0} )\frac{\vec{r}_{\textrm{Dynein}}}{|\vec{r}_{\textrm{Dynein}}|} &&  |\vec{r}_{\textrm{Dynein}}| > L_{0},
\end{align}
where $\vec{r}_{\textrm{Dynein}} = \vec{r}_{\textrm{anchor}} - \vec{r}_{\textrm{attach}}$ is the distance between the anchor and the attachment points. The measurements of elastic modulus of the stalk differ
\citep{kamiya_elastic_2016,burgess_dynein_2003,lindemann_does_2003,sakakibara_inner-arm_1999}
, therefore, the elastic modulus was estimated to $k_{\textrm{Dynein}} = 400p\textrm{N} \mu \textrm{m}^{-1}$
 \cite{sakakibara_molecular_2011}. 
If the force is null or parallel to the preferred direction of stepping, the probability of stepping to the minus end is:
\begin{equation}
p_{-} = \frac{V_{F}}{d_{\textrm{step}}},
\end{equation}
where $V_{F}$ is the forward speed of dynein and $d_{\textrm{step}}$ is the length of the step.
The steps of dynein have multiple lengths 
\cite{gennerich_force-induced_2007,toba_overlapping_2006,mallik_building_2005,mallik_cytoplasmic_2004,reck-peterson_single-molecule_2006,kural_kinesin_2005}. Nevertheless, just the most frequently measured length $d_{\textrm{step}} = 8\textrm{nm}$ is considered.
The forward speed was estimated or measured in various sources 
\cite{torisawa_autoinhibition_2014,toba_overlapping_2006,gennerich_force-induced_2007,muller_tug--war_2008,king_dynactin_2000,nishiura_single-headed_2004,kon_distinct_2004,cho_regulatory_2008,kikushima_slow_2004}.
For our purposes it is estimated to be   $V_{F} = 1000 \textrm{nm}\textrm{s}^{-1}$.
In the case of the force of the dynein being in the opposite direction to the 
preferred movement and smaller than a stall force $F_{S}$, 
the attachment point steps to the minus end with probability:
\begin{equation}
 p_{-} = \frac{V_{F}}{d_{\textrm{step}}}( 1 - \frac{|F^{\textrm{Dynein}}|}{F_{S}} ).
\end{equation}
The value of the stall force varies greatly \cite{mallik_cytoplasmic_2004,mallik_building_2005,walter_cytoplasmic_2010,gennerich_force-induced_2007,toba_overlapping_2006,muller_tug--war_2008,gennerich_force-induced_2007} and 
\cite{belyy_mammalian_2016},
we estimate it as $F_{S} = 4\textrm{pN}$.
If the force aims to the plus end and it is bigger than the stall force, dynein steps to the plus end with probability:
\begin{equation}
 p_{+} = \frac{V_{B}}{d_{\textrm{step}}}.
\end{equation}
Backward stepping speed is force dependent \cite{gennerich_force-induced_2007} 
and the measured values also differ
\cite{muller_tug--war_2008}, 
\cite{ikuta_tug--war_2014}. 
Our estimate is $V_{B} = 6.0\textrm{nm}\cdotp \textrm{s}^{-1}$  \cite{gennerich_force-induced_2007}.

\noindent
The probability of detachment is expressed as:
\begin{equation}
\label{detachment_prob}
 p_{\textrm{detach}} = \textrm{exp}(\frac{|F_{d}|}{F_{D}}),
\end{equation}
where the detachment force \cite{ikuta_tug--war_2014,muller_tug--war_2008,kunwar_mechanical_2011} was estimated as $F_{D} = F_{S} / 2$.
When the attachment point of dynein motor is not on a bead of MT, the force is acting on a point of a segment
between two beads.
In such a case, the force has to be transmitted to the two closest beads.
Since the mechanism of stepping and detachment of dynein is uncertain, we use the model for kinesin stepping \cite{klein_motility_2015}. 

Dynein plays a role in two mechanisms. 
During the cortical sliding mechanism acting in the whole IS, the MT slides on the membrane and its plus-end remains free.
MT depolimerizes in the fixed position on the membrane of the cell during capture-shrinkage mechanism acting in the center of the IS. 
Without the effects of dynein, MT detaches from fixed position.

The densities of capture-shrinkage dyneins, $\rho_{IS}$, and cortical sliding dynein $\tilde{\rho}_{IS}$ vary through the range that could be expected during the T-Cell activation,  $0 < \rho_{IS}, \tilde{\rho}_{IS} < \rho_{\textrm{MAX}}$.
The maximum density of dyneins were estimated considering its structure and size.
The dynein comprises a long stalk and a ring-like head containing six AAA+ modules whose diameter is comparable with the length of the stalk \citep{roberts_aaa_2009,imai_direct_2015}, and  N- and C-terminal regions.
The size of the dynein motor can be also estimated from the distance $d_{\textrm{hm}}$ between  the head domain of the dynein and the center of the attached MT \cite{mizuno_three-dimensional_2007}, $d_{\textrm{hm}}=28n\textrm{m}$, which substantially exceeds the length of the dynein stalk. 
For the case of simplicity, we compute the area of plasma membrane covered by one dynein as $a_{\textrm{d}} = \pi * L_{0}^{2}$, where $L_{0}$ is the length of the stalk.
The number of dynein $N_{\textrm{dynein}}$ in $1\mu\textrm{m}^{2}$ 
is calculated as  
\begin{equation}
N_{\textrm{dynein}} = \frac{1^{-12}}{a_{\textrm{d}}} \sim 1000.
\end{equation}
Consequently, $\rho_{\textrm{MAX}} = 1000\mu\textrm{m}^{-2}$.

\subsection{Microtubule organizing center (MTOC)}
\label{Model_of_the_MTOC}

\begin{figure}[ht]
     \includegraphics[trim=80 640 50 68,clip,width=0.9999\linewidth]{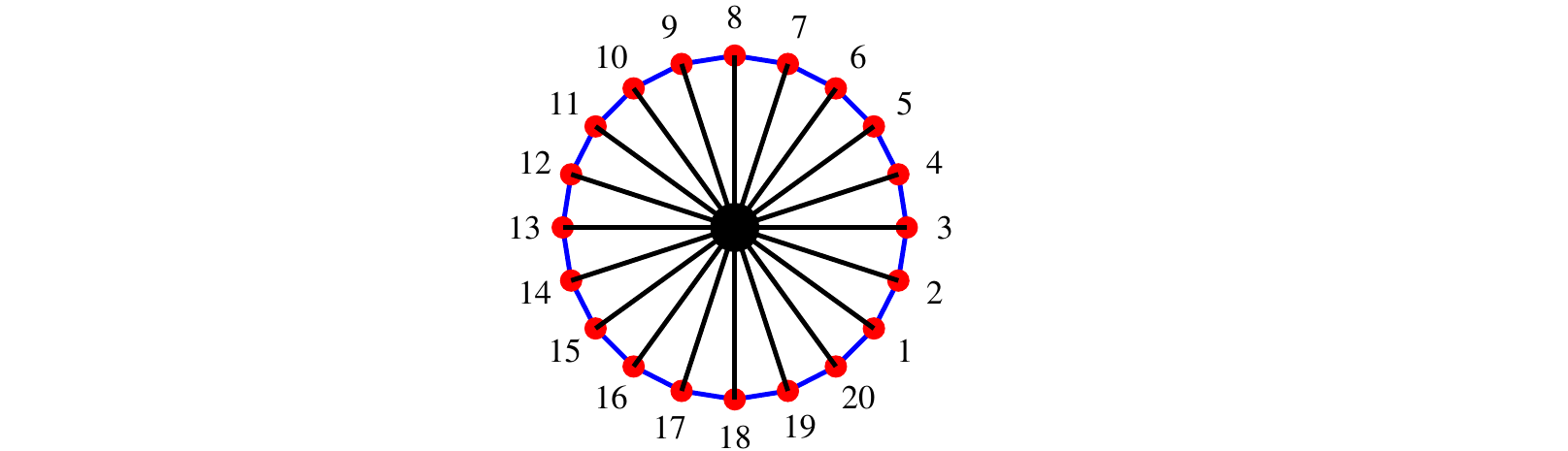}
\caption{Sketch of the MTOC with $N_{\textrm{MTOC}} = 20$ MT sprouting points represented by red spheres.
The black sphere in the middle depicts the center of the MTOC. 
Black lines connecting the points with the center and blue lines connecting the neighboring points are nondeformable.
\label{MTOC_scatch} }
\end{figure}

The MTOC is modeled as a planar, polygonal structure, composed from so-called sprouting points(points of MT sprouting).
If MTOC has $Q^{\textrm{MTOC}}$ sprouting points, then the equal number of constraints holds them
in a specified distance from MTOC center(black lines in Fig. \ref{MTOC_scatch}). 
Therefore, the $i$th constraint is defined as:
 \begin{align}
  C^{\textrm{MTOC}}_{i} = |\vec{r}^{\textrm{mtoc}}_{i} - \vec{r}_{c} | = R_{\textrm{MTOC}} && i = 1,..., Q^{\textrm{MTOC}},
 \end{align}
where $\vec{r}^{\textrm{mtoc}}_{i}$ is the position of $i$th sprouting point and $\vec{r}_{c}$ is the position of the center of the MTOC.
Moreover, additional $Q^{\textrm{MTOC}}$ bonds keep the neighboring points in a constant distance $d^{\textrm{MTOC}}$(blue lines in Fig. \ref{MTOC_scatch}).

\begin{align}
C^{\textrm{MTOC}}_{i} = |\vec{r}^{\textrm{mtoc}}_{i} - \vec{r}^{\textrm{mtoc}}_{i + 1} | = d^{\textrm{MTOC}} 
&& i =  Q^{\textrm{\textrm{MTOC}}} + 1,..., 2 \cdot Q^{\textrm{\textrm{MTOC}}}  ,
\end{align}

\begin{figure}[ht]
     \includegraphics[trim=80 630 50 68,clip,width=0.9999\linewidth]{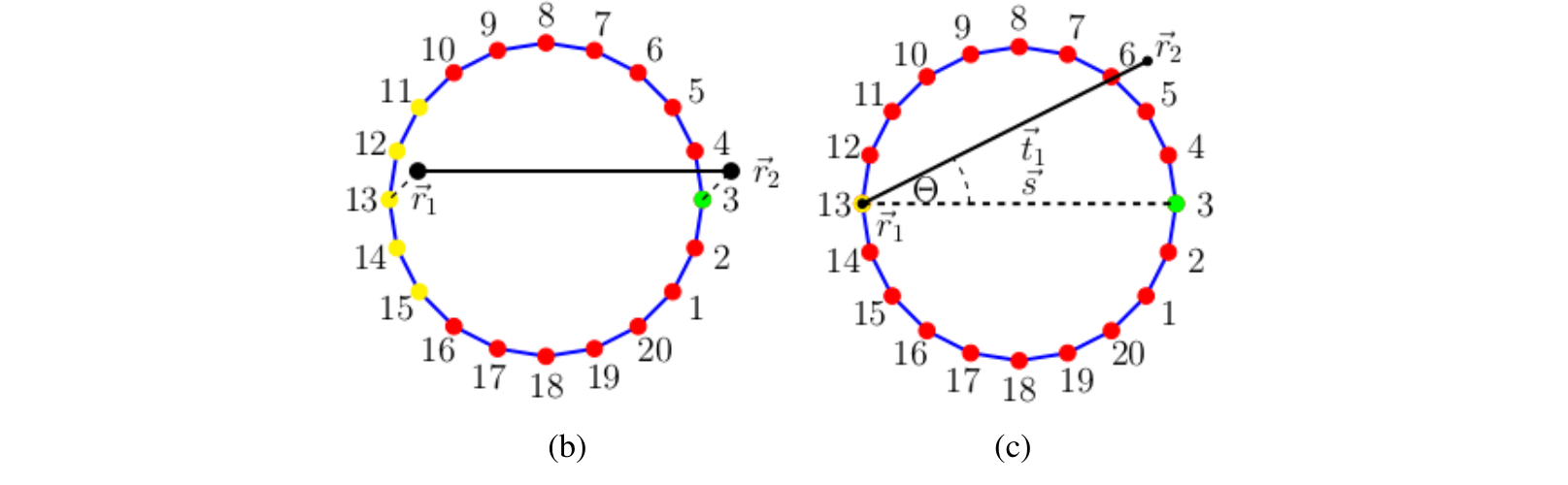}
\caption{The sketch of the MTOC and forces connecting the MTOC with MTs.
The black line denotes the first segment of microtubule. 
(a) 
Dashed lines depict the elastic forces connecting two points of microtubule 
to two MTOC points.
The green point represents the sprouting point and yellow points depict possible rear points. 
(b) Bending force minimizes the angle $\Theta$ between the first segment of microtubule $\vec{t}_{1}$ and the line $\vec{s}$ connecting sprouting(green) and rear(yellow) point. 
\label{MTOC_scatch_2} }
\end{figure}

When MT is created, the so-called "sprouting point" and "rear point" on the MTOC are chosen \ref{MTOC_scatch_2}.
The second bead of the MT is attached to the sprouting point and the first bead to the rear point.
Consequently, the original MT orientation is given by the direction from the rear point to the sprouting point. 
Every MTOC point is the sprouting point to the same number of MTs.
The rear point is chosen from the points at the approximately opposite side(relatively to sprouting point) of the MTOC \ref{MTOC_scatch_2}a, which gives a variety of the
initial MT orientations.
Elastic force \ref{MTOC_scatch_2}a anchors the MT in the MTOC, while bending force \ref{MTOC_scatch_2}b
forces the MT to be aligned with the line connecting sprouting and rear points.
The combination of two forces assures anchoring of the MT and limitation
of changes in orientation, simulating the effect of PCM.

\subsubsection{Connecting microtubule and MTOC }
\label{sec:model_of_MTOC}
The first segment of the MT is inside the MTOC \ref{MTOC_scatch_2}a. 
The second and the first the MT beads are attached by elastic forces to
the sprouting point and the rear point, respectively \ref{MTOC_scatch_2}a.
Elastic forces acting on a MT bead can be written as:
\begin{equation}
\label{micro_MTOC}
 \vec{F}^{\textrm{elas}}_{2} = k_{3} |\vec{d}_{2}| \cdot \frac{\vec{d}_{2}}{|\vec{d}_{2}|}  ,
\end{equation}
where $k_{3} = 30 \cdotp  \textrm{pN} \mu \textrm{m}^{-1} $ is a spring constant 
and $ \vec{d}_{2} = \vec{r}_{\textrm{s}}^{\textrm{ MTOC }} - \vec{r}_{2}^{\textrm{ micro }}$, 
where $\vec{r}_{\textrm{s}}^{\textrm{ MTOC }}$ and $\vec{r}_{2}^{\textrm{ micro }}$ are MTOC sprouting point and the second bead of MT,
respectively. 
Analogically, we can define the forces between the first bead of the MT and the MTOC rear point. 
Bending forces \ref{MTOC_scatch_2}b are calculated via:
\begin{subequations}
\label{mtoc_bending_forces}
\begin{align}
          \vec{F}^{\textrm{bend}}_{\textrm{MTOC}} = \frac{ \kappa }{ |\vec{s}|^{2} }\left( - \frac{\vec{t}_{1}}{|\vec{t}_{1}|}
  + \frac{\vec{s}}{|\vec{s}|} \left( \frac{\vec{t}_{1} \cdotp \vec{s} }{ |\vec{s}| |\vec{t}_{1}| } \right)  \right)\\
         \vec{F}^{\textrm{bend}}_{\textrm{micro}} = \frac{ \kappa }{ |\vec{t}_{1}|^{2} }\left(\frac{\vec{s}}{|\vec{s}|}
 - \frac{\vec{t}_{1}}{|\vec{t}_{1}|} \left( \frac{\vec{t}_{1} \cdotp \vec{s} }{ |\vec{s}| |\vec{t}_{1}| } \right)  \right),\\
 \vec{F}_{\textrm{0}} = - \vec{F}^{\textrm{bend}}_{\textrm{micro}} - \vec{F}^{\textrm{bend}}_{\textrm{MTOC}}
\end{align}
\end{subequations}
where $\vec{s}$ is the segment between the two beads of the MTOC, $\vec{t}_{1}$ is the first segment of MT.
The forces $\vec{F}^{\textrm{bend}}_{\textrm{MTOC}}$ and $\vec{F}^{\textrm{bend}}_{\textrm{micro}}$ act on the  sprouting point and the second bead of MT, respectively.
The force $ \vec{F}_{\textrm{0}}$ acts on the first bead of the MT and the rear point.

\section{\large{\bf{Constrained Langevin dynamics}}}
Using Langevin dynamics, the motion of an unconstrained particle with the position $x_{i}$ can be expressed:
\begin{equation}
\label{basic_Langevin}
\gamma_{i} \dot{x}_{i} = f_{i} + \eta_{i},
\end{equation}
where $\eta_{i}$ is a random Langevin force, which is a stochastic, non-differentiable function of time that integrates random interactions with the molecules of the solvent.
The force $f_{i}$ is the sum of all other forces and it depends on the object and  $\gamma$ is the drag coefficient.
In a constrained case,  N beads in 3D have to satisfy $Q$ constraints \cite{montesi_brownian_2005}:
\begin{align}
\label{constraints}
C_{a}( x_{1}, ..., x_{ 3* N } ) = c_{a} && a = 1,....,Q
\end{align}

\noindent
The constraints have to remain constant in every instance. 
Therefore, the movement of the beads must satisfy:
\begin{align}
\label{constraints_22}
0 = \dot{C}_{a} =  n_{i a} \cdotp \dot{x}_{i} && a = 1,....,Q 
\end{align}
where 
\begin{equation}
n_{ia} = \frac{\partial C_{a}}{\partial x_{i}}. 
\end{equation}
The motion of a constrained bead can be expressed:
\begin{equation}
\label{constraint_basic}
 \gamma \dot{x}_{i} =f_{i} + \eta_{i} - n_{ia} \lambda_{a},
\end{equation}
where $\lambda_{a}$ is the constraint force conjugate to the constraint $\mu$.

\subsection{Mid-Step algorithm }
\label{Mid-Step-algorithm}
Mid-Step algorithm was proposed by Fixman 
and further generalized by Hinch and Grassia 
\citep{fixman_simulation_1978,hinch_brownian_1994,grassia_computer_1995,grassia_computer_1996}
The algorithms was elaborated for specific cases by Morse and Pasquali 
\cite{montesi_brownian_2005} and \cite{pasquali_efficient_2002}.
Using the mobility tensor 
\begin{equation}
\label{mobility_tensor}
H_{ik} \gamma = \mathbf{I}_{ik},
\end{equation}

\noindent
the equation (\ref{constraint_basic}) can be rewritten as: 
\begin{equation}
\label{langevin_with_constraints}
 \dot{x}_{i} = H_{ij} [ F^{u}_{j} - n_{ja} \lambda_{a} ],
\end{equation}
where $F^{u}_{j} = f_{i} + \eta_{i} $ is unconstrained force. 
The values of $\lambda_{a}$ for $a = 1 , ... , Q$ can be calculated from the conditions \eqref{constraints_22}
at every instant. It will result in the set of algebraic equations:
\begin{equation}
\label{constraint_3}
 G_{a \nu} \lambda_{\nu} = n_{ia} H_{ij} F^{u}_{j},
\end{equation}
where 
\begin{equation}
 G_{a \nu} = n_{ia} H_{ij} n_{j\nu}.
\end{equation}

\noindent
If the constraint forces are expressed by (\ref{constraint_3}), we get the equation of motion from (\ref{constraint_basic}):
\begin{equation}
\label{speed_projection}
 \dot{x}_{i} = P_{ij}H_{jk}F_{k}^{u},
\end{equation}
where 
\begin{equation}
\label{dynamical_projection_operator}
 P_{ij} = \mathbf{I}_{ij} - H_{ik} n_{ka} G_{a \nu}^{-1}n_{j\nu}
\end{equation}
is a projection operator. 
In the case when the mobility tensor is expressed by (\ref{mobility_tensor}), equation \ref{dynamical_projection_operator} can be rewritten as:
\begin{equation}
  P_{ij} =  \mathbf{I}_{ij} - n_{ia} T_{a \nu}^{-1} n_{j\nu}, 
\end{equation}
where
\begin{equation}
T_{a \nu} = n_{ia} n_{i\nu}.
\end{equation}
The dynamical projection operator is used to project forces to $3N - Q$ dimensional hypersurface. 
Therefore, they are locally perpendicular to the constraints.

The mid-step algorithm proposed by Hinch is for the case of mobility tensor (\ref{mobility_tensor})
composed by four following substeps:
\begin{enumerate}
 \item Generate unprojected random forces $\eta_{i}$ and unprojected forces $f_{i}$ at initial position $x^{0}_{i}$;
 \item Construct projected random force $\eta^{P}_{i} = P_{ij} \eta_{j}$ and $f^{P}_{i} = P_{ij} f_{j}$; 
 \item Calculate midstep position $x_{i}^{1/2} = x^{0}_{i} + \dot{x}_{i}^{0} \Delta t / 2$, where the mobility
 in the original configuration $\dot{x}_{i}^{0}$
 is calculated via (\ref{speed_projection}) and  $\Delta t$ is the time step;
 \item Calculate updated bead positions $x^{1}_{i} = x^{0}_{i} + \dot{x}^{1/2}_{i} \Delta t$,
 where $\dot{x}^{1/2}_{i}$ is evaluated with the deterministic and normal vectors from mid position, but with the same projected
 random force from initial configuration.
\end{enumerate}
Mid-step algorithm uses the projection operator \eqref{dynamical_projection_operator} that alongside with the mid position calculation minimizes the perturbations of constraints. 
Nevertheless, perturbations cannot be eliminated.
Therefore, the MT has to be resized to fulfill the constraints.
In such operation, the angles between $\vec{t}_{i}$ and $\vec{t}_{i+1}$, where $i = 1,...,N-1$ are conserved, the first bead of MT remains constant and MT regrows from the MTOC.
Consequently, the bending energy of the MT remains unchanged.

\section{Additional results}
\label{Additional_results}

\subsection{Influence of random forces}
\label{Influence_of_random_forces}
\begin{figure*}[ht]
     \includegraphics[trim=80 530 50 68,clip,width=\linewidth]{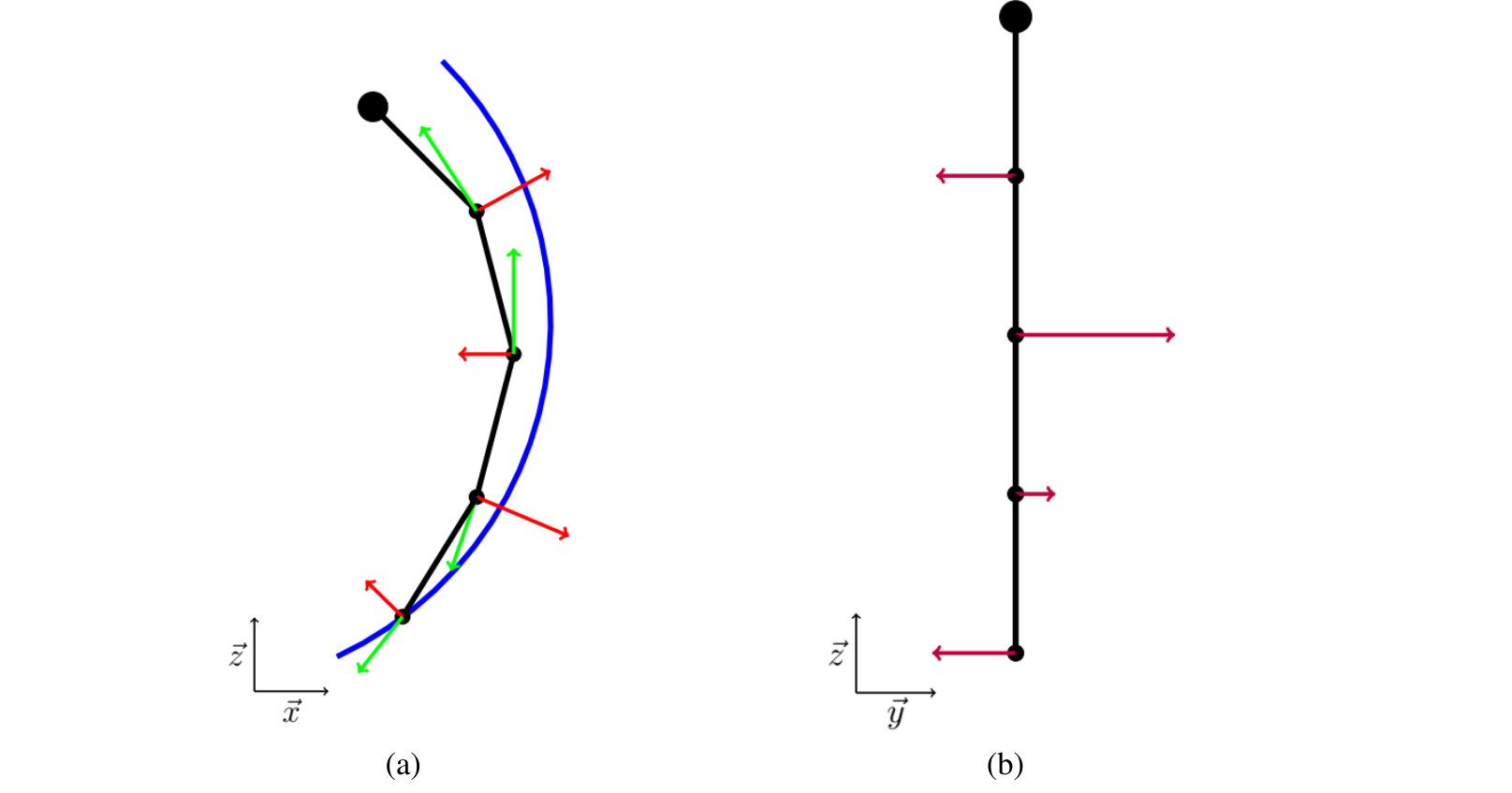}
   \caption{Sketch of random forces acting on the MT from two perspectives. 
   The big circle represents MTOC and smaller circles represent beads of microtubule. 
Blue arc depicts membrane of the cell.
Red, green and purple lines representing random forces acting on every bead are perpendicular to each.   
   \label{fig:microtubule_random_scatch}}   
\end{figure*}

\begin{figure*}[ht]
     \includegraphics[trim=40 380 50 68,clip,width=0.66\linewidth]{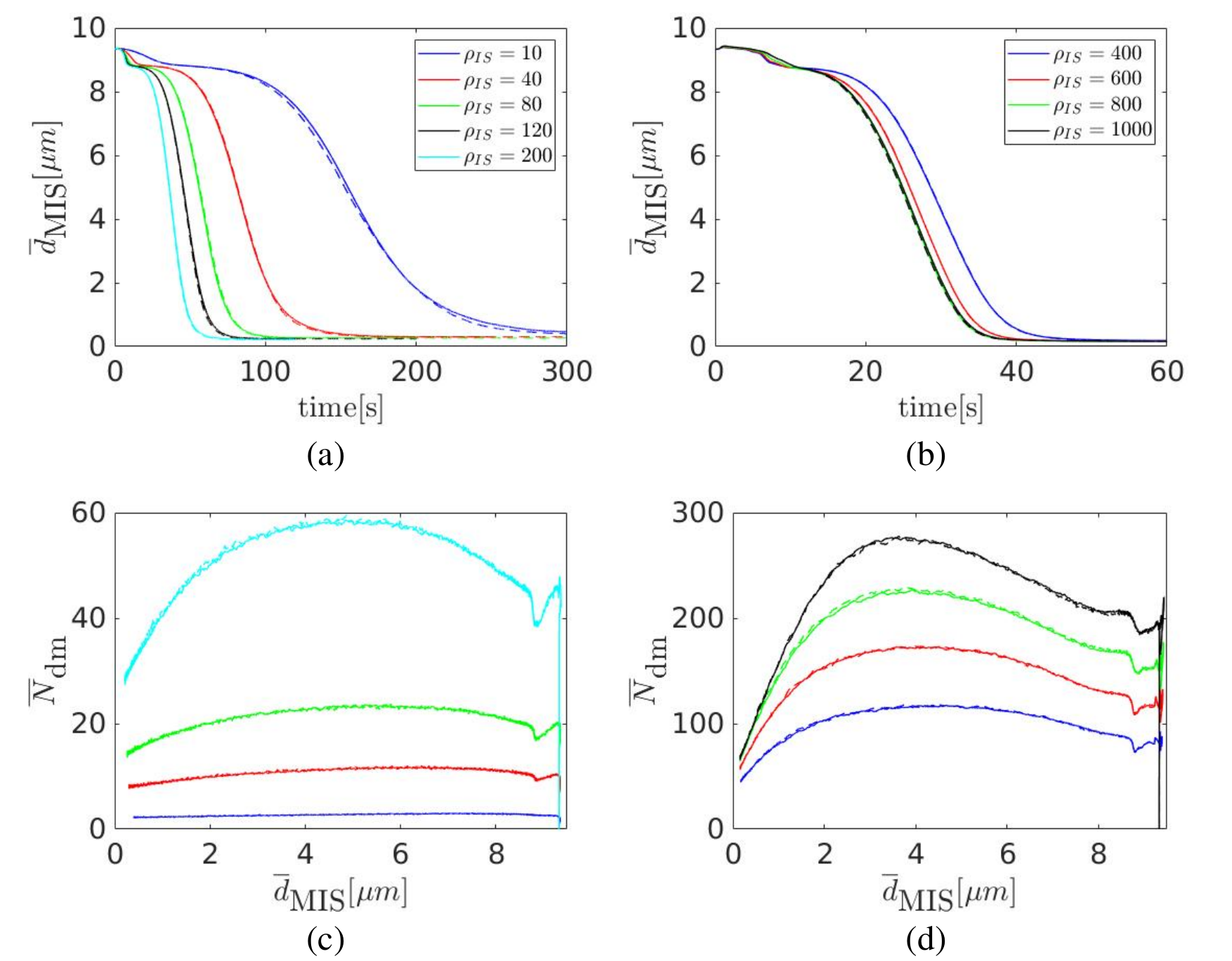}
    \caption{Combination of the capture-shrinkage mechanism and random forces. Solid lines stand for the sole capture shrinkage and dashed lines for the combination with random forces. Legends in (a), (b) apply for (c), (d), respectively.
 (a)(b) Dependence of the average MTOC-IS  distance $\bar{d}_{\textrm{MIS}}$ on time, (c)(d) Dependence of the average number of dynein
$\bar{N}_{\textrm{dm}}$    
     motors on MTOC-IS distance $\bar{d}_{\textrm{MIS}}$. \label{fig:capture_shrinkage_random}} 
 \end{figure*}

\begin{figure*}[ht]
     \includegraphics[trim=40 550 50 68,clip,width=0.66\linewidth]{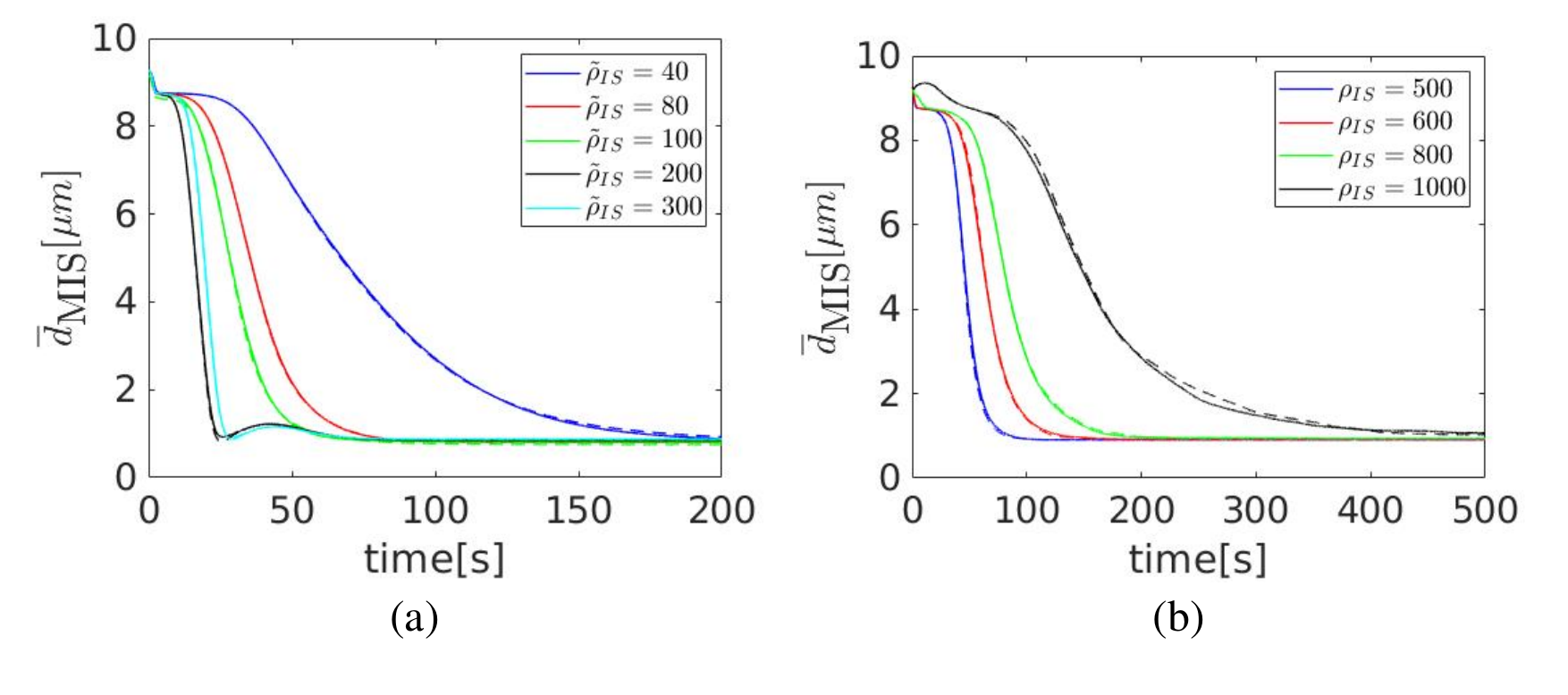}
   \caption{Combination of the cortical sliding mechanism and random forces. 
   Solid lines stand for the sole cortical sliding and dashed lines for the combination with random forces.
(a)(b) Dependence of the average MTOC-IS distance $\bar{d}_{\textrm{MIS}}$ on time.
   \label{fig:cortical_shrinkage_random}} 
 \end{figure*}

Random forces acting on the MTs have small effects since the motion of the MTs is constrained.
Fig. \ref{fig:microtubule_random_scatch}a suggests that random forces perpendicular to the cell membrane(red) have no effect since
microtubule curvature is given by the interplay of bending forces and the force of the membrane. 
The green forces, acting parallel to the  segments  of the MTs 
have also no impact, because the MT is attached to the MTOC, making it a part of a very massive structure.
Consequently, just random forces depicted in purple in \ref{fig:microtubule_random_scatch}b can result in movement, making the random noise effectively one-dimensional.
However, the MT is still rigid structure, random forces act in contradiction and filament is bound to MTOC, which seriously limits the movement of upper beads.
Therefore, the influence of random noise can be expected to be negligible.  
Moreover, the capture-shrinkage mechanism fixates the MT on both sides, further minimizing the effect of the random force.
In Figs. \ref{fig:capture_shrinkage_random}a and \ref{fig:capture_shrinkage_random}b we can see that the repositioning curves 
are almost identical for the case of the capture-shrinkage mechanism.
The Figs. \ref{fig:capture_shrinkage_random}c and \ref{fig:capture_shrinkage_random}d demonstrate that the developments of the number of attached dyneins also do not differ. 
During cortical sliding repositionings  the effect of the random forces is also negligible \ref{fig:cortical_shrinkage_random}.

\subsection{Comparison of two cases with different number of microtubules}
\begin{figure*}[ht]
     \includegraphics[trim=40 370 50 68,clip,width=0.66\linewidth]{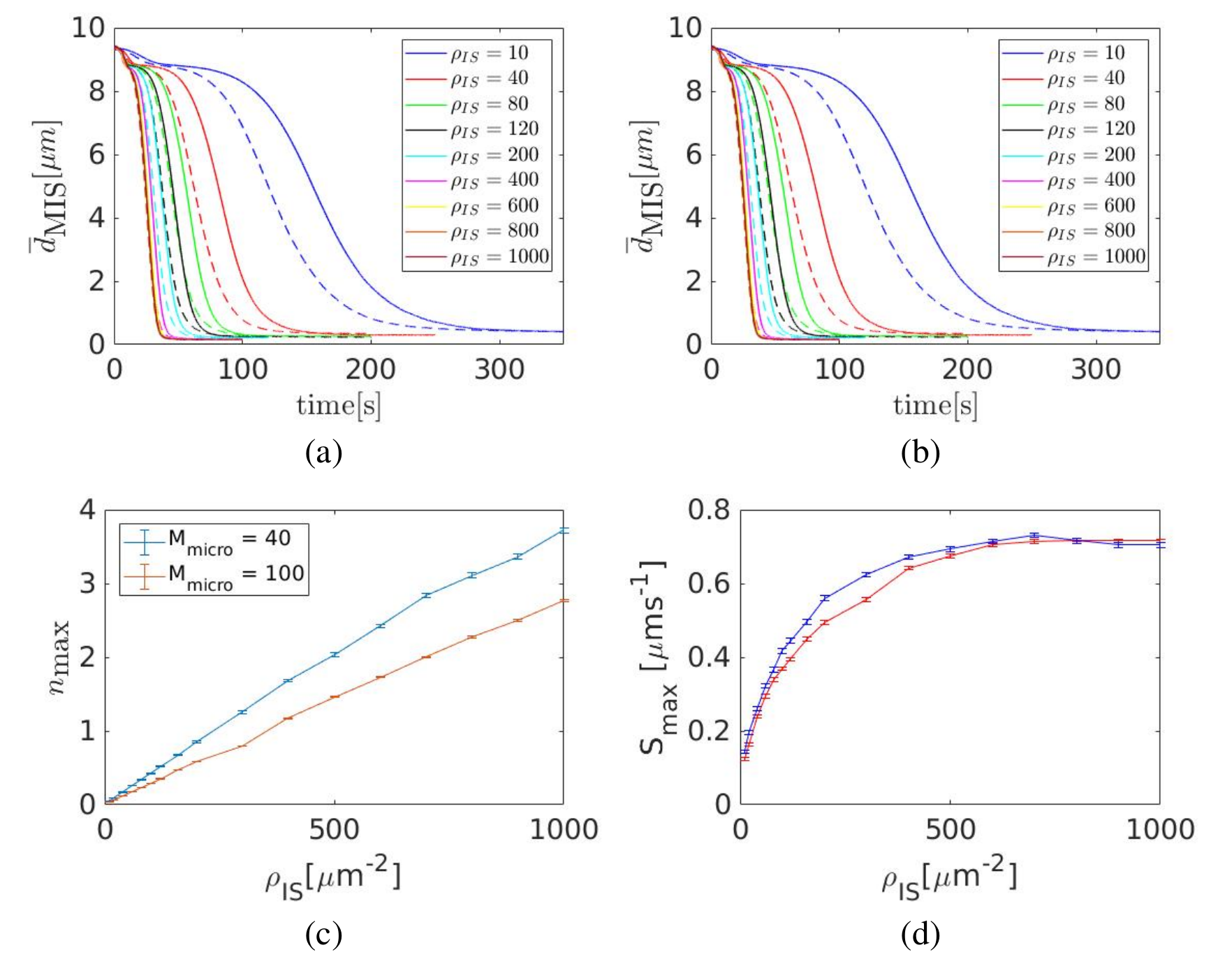}
   \caption{Capture-shrinkage mechanism for two cytoskeletons with different numbers of microtubules: $M_{\textrm{micro}} = 100$(solid lines) and $M_{\textrm{micro}} = 40$(dashed lines).
(a) Dependence of average MTOC-IS distance $\bar{d}_{\textrm{MIS}}$ on time.
(b) Dependence of average MTOC speed $\bar{v}_{\textrm{MTOC}}$ on $\bar{d}_{\textrm{MIS}}$. 
(c) Dependence of the maximum number of attached dyneins per microtubule $n_{\textrm{max}}$ on area density $\rho_{\textrm{IS}}$. 
(d) Dependence of  the maximum speed $S_{\textrm{max}}$ on $\rho_{\textrm{IS}}$.
       \label{fig:Capt_sh_comparison} } 
 \end{figure*}

The cytoskeleton of $M_{\textrm{micro}} = 100$ MTs (examined in previous section) is compared with the cytoskeleton of $M_{\textrm{micro}} = 40$. 
We define $n_{\textrm{dm}}(t) = \frac{\bar{N}_{\textrm{dm}}(t)}{M_{\textrm{micro}}}$ to examine the ratio of the attached dyneins and the number of the MTs in a cytoskeleton.
Fig. \ref{fig:Capt_sh_comparison}a depicts repolarization curves of two cytoskeletons for the case of the capture-shrinkage mechanism.
The polarization exhibits a triphasic behavior for both cases  \ref{fig:Capt_sh_comparison}b.

We define $n_{\textrm{max}} = \textrm{max}(n_{\textrm{dm}}(t))$.
In Figs \ref{fig:Capt_sh_comparison}c it can be seen that 
$n_{\textrm{max}}$ is always bigger for the case of smaller cytoskeleton(caused by the small area of the center of IS and limited number of dynein).
The Figs. \ref{fig:Capt_sh_comparison}c and \ref{fig:Capt_sh_comparison}d explain the differences of speed in terms of the number of motors.
When the area density is small, the smaller cytoskeleton is pulled with relatively higher force.
As the concentration increases, the maximum speed is achieved($\rho_{\textrm{IS}} \sim 600 \mu\textrm{m}^{-2}$).
Subsequent increase of pulling force has no effect.

\begin{figure*}[ht]
     \includegraphics[trim=40 380 50 68,clip,width=0.66\linewidth]{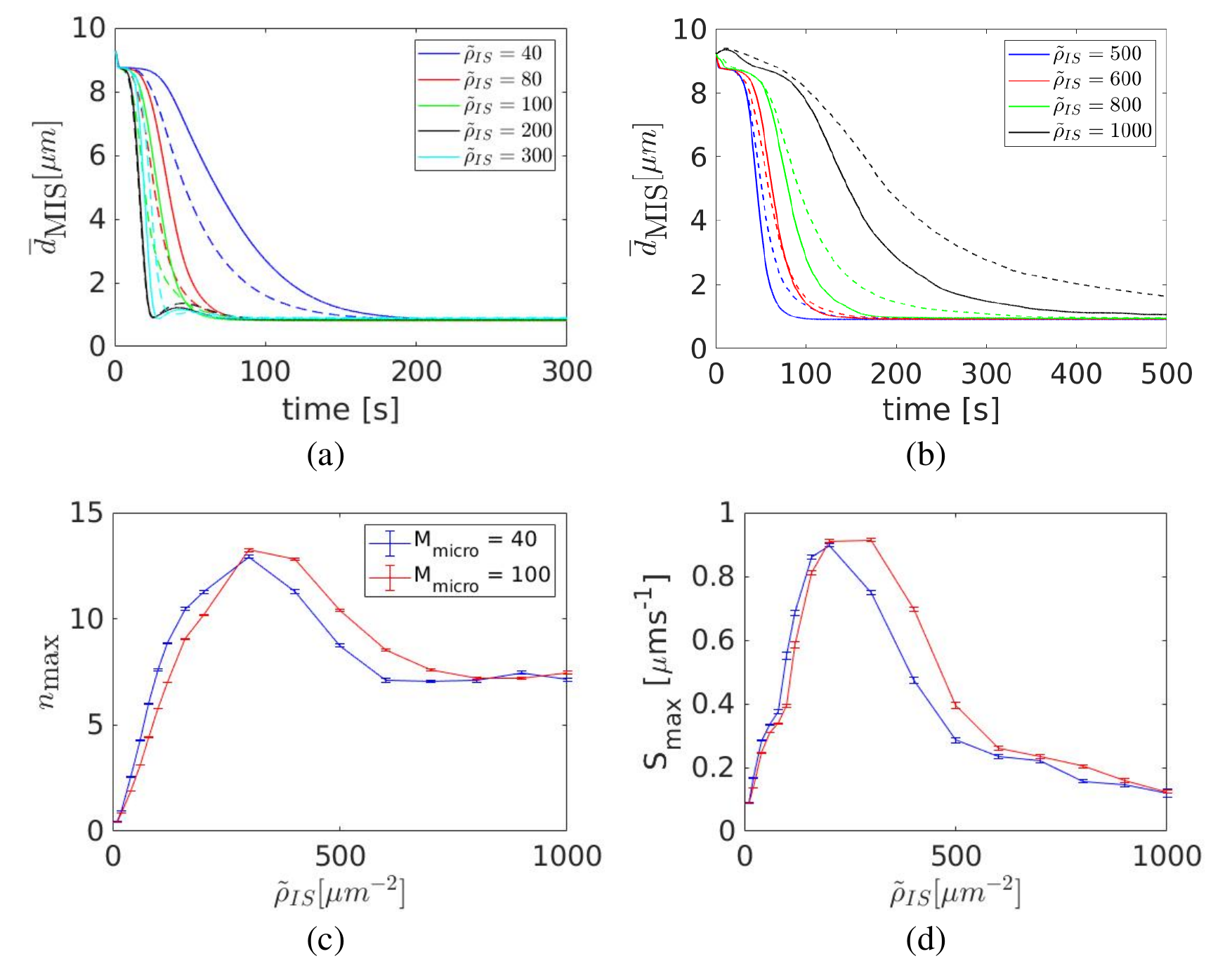}
   \caption{Cortical sliding mechanism for two cytoskeletons with different numbers of microtubules $M_{\textrm{micro}} = 100$(solid lines) and $M_{\textrm{micro}} = 40$(dashed lines).
   (a)(b) Dependence of average MTOC-IS distance $\bar{d}_{\textrm{MIS}}$ on time.
(c) Dependence of the number of attached dyneins per microtubule $n_{\textrm{max}}$ on the area density $\tilde{\rho}_{\textrm{IS}}$. 
(d) Dependence of the maximum speed $S_{\textrm{max}}$ on $\tilde{\rho}_{\textrm{IS}}$.
       \label{fig:Cortical_Sliding_comparison} } 
 \end{figure*}

Figs. \ref{fig:Cortical_Sliding_comparison}a and
 \ref{fig:Cortical_Sliding_comparison}b depicting the repositioning under the influence of the cortical sliding mechanism shows the three regimes for both cytoskeletons.
In both cases, $n_{\textrm{max}}$ rises at the beginning, it reaches its maximum when $\tilde{\rho}_{\textrm{IS}}\sim200\mu\textrm{m}^{-2}$ and then it decreases swiftly and then steadily \ref{fig:Cortical_Sliding_comparison}c.  
For smaller densities, the number of dyneins per MT are smaller for the bigger cytoskeleton.
The situation is opposite when considering high area densities.
Since the attached MTs aim in different directions, dyneins compete in the area of higher densities.
As the  number of the MT decreases, the pulling forces acting on individual filaments increase, leading to faster detachment.
The MTOC speed increases when $\tilde{\rho}_{\textrm{IS}}<200\mu\textrm{m}^{-2}$ and then it decreases.
We can see in \ref{fig:Cortical_Sliding_comparison}d that the speed decreases for both cases even when $n_{\textrm{max}}$ stays approximately the same, which is the consequence of dynein acting predominantly at the periphery.

\subsection{Capture-shrinkage and cortical sliding combined}
\begin{figure*}[ht]
     \includegraphics[trim=50 640 50 68,clip,width=0.85\linewidth]{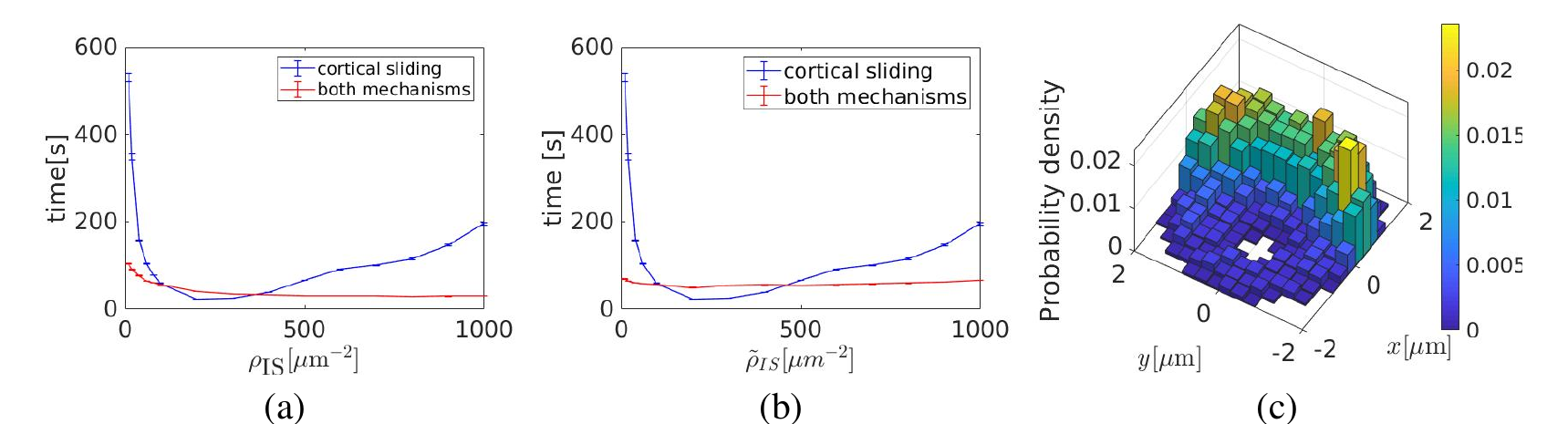}
   \caption{Combination of the capture-shrinkage and the cortical sliding mechanisms. (a)(b) Times of repositioning for the sole cortical sliding and combinations of two mechanisms.
(a)In the combined case the capture-shrinkage area density varies, cortical sliding area density is
constant $\tilde{\rho}_{\textrm{IS}} = 20\mu\textrm{m}^{-2}$.
(b)In the combined case the cortical sliding area density $\tilde{\rho}_{\textrm{IS}}$ varies and  capture-shrinkage area density is constant 
$\rho_{\textrm{IS}} = 60\mu\textrm{m}^{-2}$.
(c) Two dimensional probability density of attached dynein 
$\tilde{\rho}_{\textrm{IS}} = 60\mu\textrm{m}^{-2}$, $\rho_{\textrm{IS}} = 20\mu\textrm{m}^{-2}$, $\bar{d}_{MIS} = 4.5\mu\textrm{m}$.
    \label{fig:cortical_with_small_capture} } 
 \end{figure*}

As can be seen in Fig. \ref{fig:cortical_with_small_capture}b,
addition of the small area density of capture-shrinkage dyneins in the center of IS causes substantial decrease of differences between times of polarization.
Moreover, the three regimes of behavior based on the area density of  
cortical sliding dyneins is not observed in the presence of the capture-shrinkage mechanism.
Surely, the third regime presents a disadvantage, since the pulling force of dynein is wasted in unproductive competitions.
Therefore, the synergy of two mechanisms proves once more to be highly effective, since it does not only removes the third regime, but also greatly reduces the times of repositioning when the area density of cortical sliding 
$\tilde{\rho}_{\textrm{IS}}<100\mu\textrm{m}^{-2}$.
Fig. \ref{fig:cortical_with_small_capture}a depicts times of repositioning 
for different sets of combined mechanisms since the capture-shrinkage area density varies and cortical sliding density remains constant.
We can see that the times of repositioning are in general shorter for the case of combined mechanisms.
Moreover, even when the area densities correspond to the second regime, the times of repositioning are comparable.
Combined cases, however, have just $15\%$ of the number of dyneins.
Additionally, we can notice that the increase of area densities when $\tilde{\rho}_{\textrm{IS}}>500\mu\textrm{m}^{-2}$ presents no advantage since
it causes slowing down of repositioning in the absence of capture-shrinkage and has no effect when the mechanisms are combined. 
Fig. \ref{fig:cortical_with_small_capture}c shows that the attached dyneins are predominantly located on the periphery of IS even in the case when 
$\tilde{\rho}_{\textrm{IS}}>\rho_{\textrm{IS}}$.

\section{Commentary on modeling approaches}

\subsection{Cytosim}
Cytosim is widely accepted as an efficient tool for the simulations of fibers \cite{nedelec_collective_2007}.
Although there are many similarities between the models, we decided not to use Cytosim.
The first reason is our goal to examine the role of Brownian motion.
The implicit integration used by Cytosim has a numerical error that could influence the precision of calculation in the absence of thermal noise.
The second reason is the simplified calculation of the bending forces used by Cytosim.
The advantage of such an approach, which enables to express the bending forces as a result of a matrix-vector multiplication, is a great efficiency of calculation.
Nevertheless, the procedure is valid only if the angles between subsequent 
segments remain small.
This presents a drawback, since the angles between the segment increase
as the radius of the cell decrease.
More importantly, substantial curvature of the MTs can be expected during repositioning \cite{kuhn_dynamic_2002}(See 4.4).
Moreover, the "reshaping" of the objects due to the numerical impressions is done to keep the center of mass constant.
Since the rigidity of MTOC is an important part of our model, reshaping is done to keep the first the bead of the MT(therefore MTOC-MT forces) constant.

\subsection{Model using deterministic force}

Kim and Maly \citep{kim_deterministic_2009} modeled the cortical sliding mechanism using the deterministic force.
Although this model has various merits,
it leads to the contradiction with some experimental observable: for example MTs stalk going through the center
\citep{kim_deterministic_2009}.
This presents a drawback since various biological functions depend on the distribution of MTs.







